\documentclass[10pt,journal,compsoc]{IEEEtran}

\ifCLASSOPTIONcompsoc

  \usepackage[nocompress]{cite}
\else

  \usepackage{cite}
\fi

\ifCLASSINFOpdf

\else

\fi
\usepackage{times,amsmath,epsfig}
\usepackage{fancyhdr}
\usepackage{amsfonts}
\usepackage{amssymb}
\usepackage[latin1]{inputenc}
\usepackage{array}
\usepackage{graphicx}
\usepackage{url}
\usepackage{subfig}
\usepackage{bm}
\usepackage{breqn}
\usepackage{xcolor}
\usepackage{soul}
\usepackage{palatino}
\usepackage{tikz}
\usepackage{multirow}
\usepackage{comment}
\usepackage{upgreek}
\usetikzlibrary{shapes.geometric, arrows}
\newcommand{\ud}{\mathrm{d}}

\graphicspath{{figures/}}

\begin{document}
\bibliographystyle{ieeetr}
%
\title{Field Distortion Model Based on Fredholm Integral}
%
%
%
%

\author{Yunqi~Sun,
        Jianfeng~Zhou

\IEEEcompsocitemizethanks{\IEEEcompsocthanksitem Jianfeng Zhou is with the Department of Engineering Physics and Center for Astrophysics, Tsinghua University, Beijing, 100084, China and Xingfan Information Technology Co., Ltd.(Ningbo), Zhejiang 315500, China. E-mail: jianfeng\_zhou@qq.com 

\IEEEcompsocthanksitem Yunqi Sun is with the Center for Astrophysics and Department of Physics, Tsinghua University, Beijing, 100084, China. E-mail: yq-sun17@mails.tsinghua.edu.cn}

\thanks{Manuscript received January 21, 2021; revised August 26, 2021. This work was supported in part by the National Program on Key Research and Development Project (Grant No. 2016YFA0400802), and in part by the National Natural Science Foundation of China (NSFC) under Grant No. 11173038 and No. 11373025.
}}

%
%

\markboth{Journal of Pattern Analysis and Machine Intellengience,~Vol.~14, No.~8, August~2015}%
{Sun \MakeLowercase{\textit{et al.}}: Field Distortion Model Based on Fredholm Integral}
%



\IEEEtitleabstractindextext{%
\begin{abstract}
Field distortion is widespread in imaging systems. If it cannot be measured and corrected well, it will affect the accuracy of photogrammetry. To this end, we proposed a general field distortion model based on Fredholm integration, which uses a reconstructed high-resolution reference point spread function (PSF) and two sets of 4-variable polynomials to describe an imaging system. The model includes the point-to-point positional distortion from the object space to the image space and the deformation of the PSF so that we can measure an actual field distortion with arbitrary accuracy. We also derived the formula required for correcting the sampling effect of the image sensor. Through numerical simulation, we verify the effectiveness of the model and reconstruction algorithm. This model will have potential application in high-precision image calibration, photogrammetry and astrometry.
\end{abstract}

}

\maketitle

\IEEEdisplaynontitleabstractindextext

%
\IEEEpeerreviewmaketitle

\ifCLASSOPTIONcompsoc
\IEEEraisesectionheading{\section{Introduction}\label{sec:introduction}}
\else
\section{Introduction}
\label{sec:introduction}
\fi

%
%
%
%
\IEEEPARstart{F}{ield} distortion is a fundamental problem in imaging systems. It is usually considered as the deviations of a real imaging device from the ideal pinhole or projective camera model \cite{stein1997lens}. Field distortion causes a loss of image quality and inadequate measurement accuracy. Recent researchers have investigated models for describing and correcting field distortions to eliminate its influence in high accuracy image measurement. Classical models for field distortion are geometric models based on the pinhole imaging process. Magill proposed a changing focus model for field distortion in 1955, which Brown improved in 1971 \cite{duane1971close}. Following this basic changing focus model, Weng proposed a complete field distortion model including radial, decentering, and thin prism distortion parts for cameras \cite{weng1992camera}. Moreover, Wang presented a geometric distortion model with rotation \cite{wang_2008}.

There are several correction methods for field distortion based on these distortion models. Wang proposed a comprehensive correction method by dividing the conventional lens distortion as radial, and tangential components \cite{wang_2008}. The rational function interpolation method \cite{huang2016} can be applied to correct the distortions in the satellite photographs to the best position accuracy of 0.1 pixels. These methods usually need calibrators or control points to guarantee the correction's accuracy, but several self-calibration models could achieve high accuracy corrections without calibrators. For example, Sawhney and Kumar presented an iterative method to correct lens distortions in videos without knowing a non-distorted reference image. RMS error of pixels for the optical centre can be reduced to 0.3 pixels \cite{sawhney1999true}. Sun \cite{sun2016camera} proposed a flexible self-calibration method with optimal intrinsic parameters, which reduces the relative error to 1\% for the calibrated model's coordinates. Goljan and Fridrich \cite{Goljan_2014} provided a self-calibration method with just a single image and several prior knowledge of the camera. Fitzgibbon studied field distortion by between-view relations and proved a distortion model with a corresponding correction method that incorporates lens distortion to solve the linear estimation of two-view point correspondences \cite{fitz2001}.

Most field distortion models and correction methods assume that the imaging system follows a pinhole camera model. However, the pinhole model is vulnerable in describing general imaging systems because of its strict restrictions for the size of point spread functions(PSF). Imaging systems with PSFs larger than one pixel of those image sensors exist in many research areas. For example, X-ray and optical telescopes \cite{Moretti_2005}\cite{blakeslee2002automatic}, medical imaging systems such as Magnetic Resonance Imaging(MRI), and Computed Tomography(CT). Even if the PSF of an imaging system is smaller than a pixel, the pinhole model remains only an applicable low-resolution model unable to cover the imaging system's high-frequency information. Therefore, it is necessary to provide a general imaging model to describe the field distortion with high resolution and accuracy.

This paper offered a general field distortion model based on the Fredholm integral. The Fredholm integral can describe the imaging process as an integral of the sources and the kernel function, which is more similar to the actual imaging process, and is more suitable for high accuracy simulation and measurement for an arbitrary field distortion. Firstly, we introduced the definition of the Fredholm Integral model for field distortion in Section \ref{sec2}. In Section \ref{sec3}, we discussed a 2-step measurement method for simulated image distortions, including a correction method for CCD sampling effect and an optimal method for finding the Fredholm Integral model's polynomial parameters. Section \ref{sec4} includes the results for the measurement method we discussed in Section \ref{sec3}. Section \ref{sec5} contains the potential applications of the Fredholm Integral model and further improvements, and in Section \ref{sec6}, we presented a brief conclusion for this paper.

\section{General field distortion model}
\label{sec2}
\subsection{Represent an imaging system by Fredholm Integral}
Suppose that we have a 2D object $O(x,y)$, a projection from a 3D object to the object plane, and an observed image $I(u,v)$ on the image plane. We claim that a general imaging process could be described by a Fredholm integral, such that,
\begin{equation}
\label{eqn:fredholm}
	I(u,v) = \iint_{\mathbb{R}^2} K(u,v,x,y)O(x,y)\ud x\ud y+n(x,y),
\end{equation}
where $K(u,v,x,y)$ represents the kernel function of the imaging system, $n(x,y)$ is the noise. $(u,v)$ and $(x,y)$ represent the coordinates on image and object plane respectively. Considering the conservation of flux, the kernel function $K(u,v,x,y)$ should follow $\iint_{\mathbb{R}^2} K(u,v,x,y)\ud u\ud v =1$ for any given $(x,y)$. The origins of coordinate in both the object plane and the image plane are the projecting points of the main axis. We temporarily dismiss the influence of noise $n(u,v)$, and focus on the kernel function $K(u,v,x,y)$.

If the imaging system is shift-invariant, then the kernel $K(u,v,x,y)$ degenerates to a 2-dimensional function denoted by $p(u-x, v-y)$, which represents the PSF of the imaging system. Thus, the imaging process could be described by
\begin{equation}
\label{eqn:nondistorted}
	I(u,v) = \iint_{\mathbb{R}^2} p(u-x,v-y)O(x,y)\ud x\ud y.
\end{equation}

High accuracy models for a real imaging system is commonly shift-variant, and it can be well approximated by a shift-invariant imaging model with lower accuracy. Therefore, we assume that PSFs at different positions in a real imaging system are similar. If we choose an arbitrary given PSF as a reference PSF $p_d(u,v,x_0,y_0)$ for the point source at $(x_0,y_0)$, which can be noted as $p_d(u,v)$, other PSFs can be obtained by adding distortion functions to the reference PSF. Then the imaging system could be modeled by,

\begin{equation}
\label{eqn:distorted}
	I(u,v) = \iint_{\mathbb{R}^2} p_d(u-x+f,v-y+g)O(x,y)\ud x\ud y,
\end{equation}
where  $f(u,v,x,y)$ and $g(u,v,x,y)$ are distortion functions. To ensure the shape of reference PSF is not affected by the distortion functions,  $f,g$ should follow that $\lim_{x\to x_0,y\to y_0} f(u,v,x,y) = 0$ and $\lim_{x\to x_0,y\to y_0} g(u,v,x,y) = 0$, which is inferred as distortion free condition(DFC) in the following context. 

\subsection{Polynomial approximation of the distortion functions}
We apply polynomial approximation to estimate the complicated distortion functions $f(u,v,x,y)$ and $g(u,v,x,y)$ to any given accuracy, such that,
\begin{eqnarray}
\label{eqn_poly}
	f(u,v,x,y) = \sum_{i,j,m,n}a_{ijmn}x^iy^ju^mv^n\\
	g(u,v,x,y) = \sum_{i,j,m,n}b_{ijmn}x^iy^ju^mv^n,
\end{eqnarray}
where $i,j,m,n = 0,1,2,...$; $i+j\geq1$. These constraints are given by the DFC.

These distortion functions are shift-variant intrinsically if $f,g$ are none trivial. Note that for two different identical point sources at $(x_1,y_1)$ and $(x_2,y_2)$, their responses on the image plane are $p_1(u,v) = p_d(u-x_1+f(u,v,x_1,y_1),v-y_1+g(u,v,x_1,y_1))$ and $p_2(u,v) = p_d(u-x_2+f(u,v,x_2,y_2),v-y_2+g(u,v,x_2,y_2))$. $p_1(u,v)$ and $p_2(u,v)$ are normally unequal since $(x_1,y_1)$ and $(x_2,y_2)$ are different. 

\subsection{Simulations of the fredholm integral model}
\label{subsec_simu4model}
We present simulations for field distortion by the Fredholm Integral model with several types of distortion functions, including polynomial functions and logarithmic functions. We choose a standard gaussian reference PSF $p(u,v)$ defined as,
\begin{equation}
	\begin{aligned}
	p(u,v)=\frac{1}{2\pi\sigma^2}\exp(-\frac{u^2+v^2}{2\sigma^2}).
	\end{aligned}
\end{equation}
By attaching the distortion functions to the reference PSF, we obtain the following distortion kernel for Fredholm Integral model,
\begin{equation*}
	K_d(u,v,x,y)=\frac{1}{2\pi\sigma^2}\exp(-\frac{(u-x+f)^2+(v-y+g)^2}{2\sigma^2}).
\end{equation*}

The 2D object for simulation is a $9\times 9$ point source array. Every single point source have a photon count of $10^5$. It is demonstrated on a $505$ pixel $\times505$ pixel image shown in Fig. \ref{fig:source}. The coordinate system of the following figures has the same unit for demonstration and simulation, while their origins are different. For calculating the simulated distortions, we assume that the origin of the figure is at its center pixel. For demonstrating the simulations, we show them in a natural image coordinate system where the pixels increase from top-left to bottom-right.

\begin{figure}[!t]
	\centering
	\includegraphics[scale = 0.5]{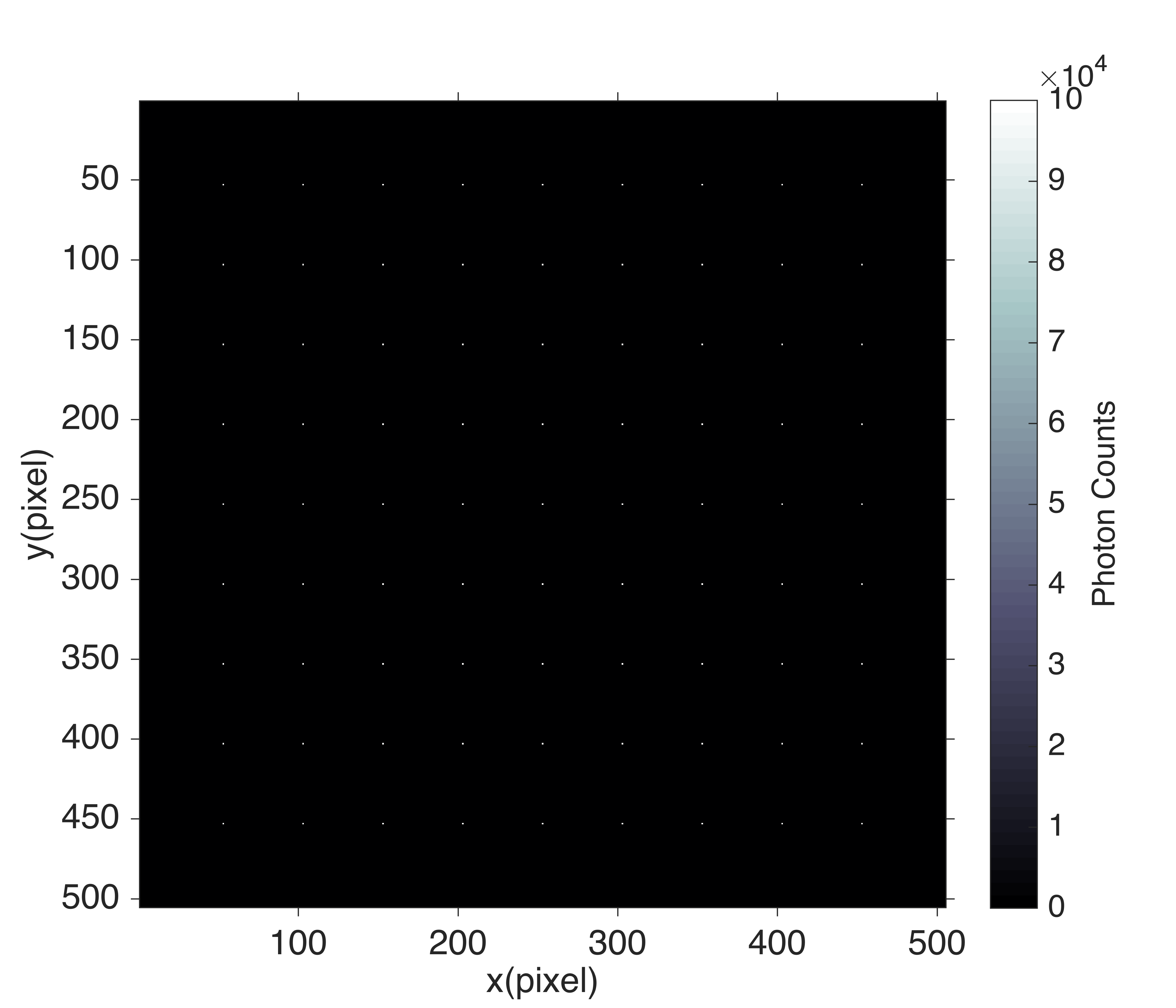}
	\caption{2D object for simulation. This input image has a $9\times 9$ point source array evenly distributed over a $505$ pixel$\times 505$ pixel size. The center of the image is defined to be $(0,0)$ and the distance between each pair of adjacent point sources are 50 pixels.}
	\label{fig:source}
\end{figure}

\begin{figure*}[!t]
	\centering
	\subfloat[]{\includegraphics[width=2in]{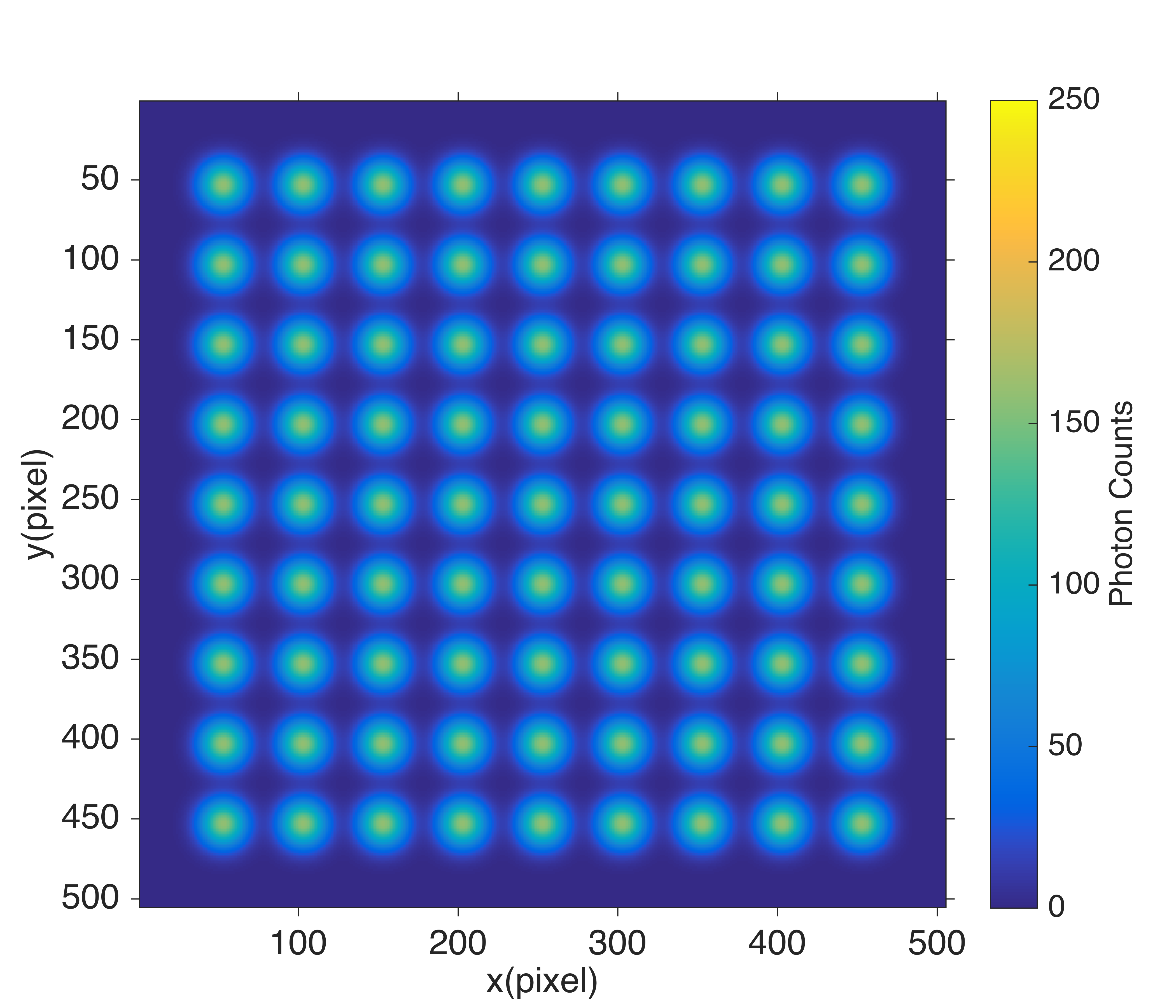}
	\label{fig_first_case}}
	\hfil
	\subfloat[]{\includegraphics[width=2in]{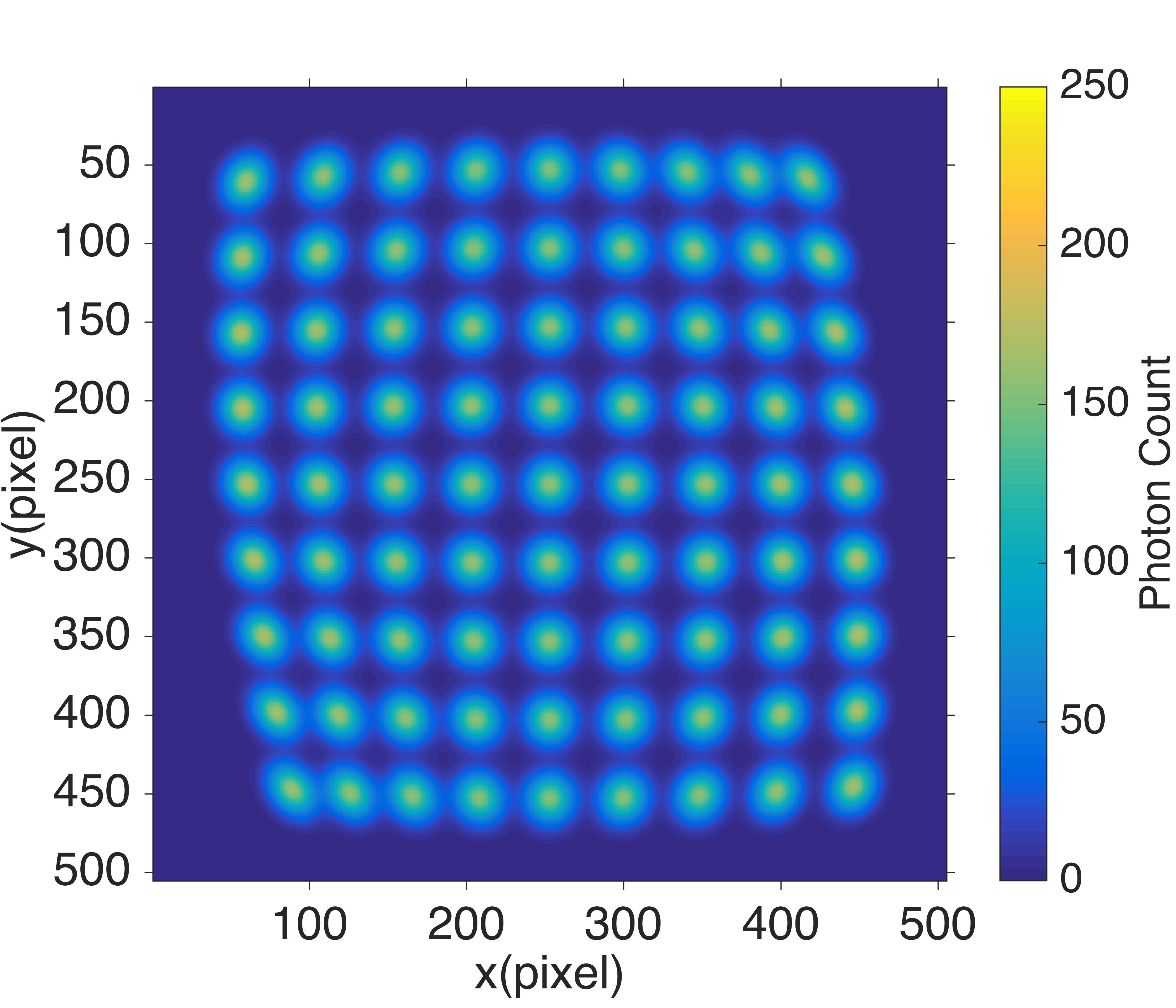}
	\label{fig_multipoly}}
	\hfil
	\subfloat[]{\includegraphics[width=2in]{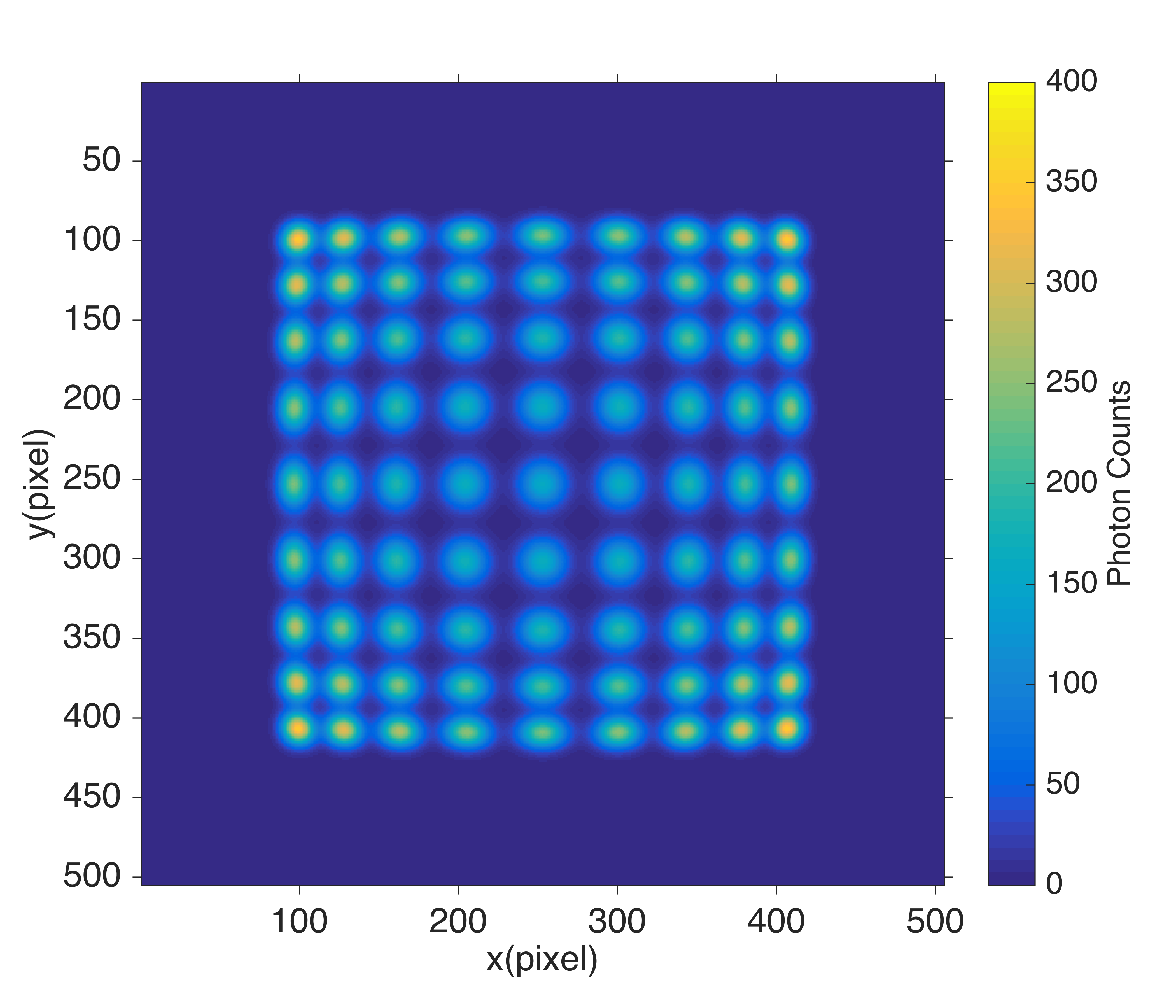}
	\label{fig_log}}
	\caption{Simulated distortions with Fredholm Integral Model with Gaussian reference PSF($\sigma = 10$). (a) is a simulation for the shift-invariant kernel function with $f(u,v,x,y) = g(u,v,x,y) = 0$. (b) is a simulation for polynomial distortion functions $f(u,v,x,y) = 10^{-6}\times(2vx+u^2x+2v^2x-2vx^2)$, $g(u,v,x,y) = 10^{-6}\times(3vy+u^2y-v^2y+vy^2)$. (c) is simulations for a logarithmic distortion function, such that $f(u,v,x,y)=x\log(1+10^{-5}(u^2+v^2))$,$g(u,v,x,y)=y\log(1+10^{-5}(u^2+v^2))$.}
	\label{fig_example}
\end{figure*}

We simulate the distortion model in the following strategy. First, we input the object and a distortion kernel to the Fredholm Integral model to calculate an observation. Then the observation is sampled and recorded by the $505\times 505$ detector array. This simulation includes both polynomial distortion functions and logarithmic distortion functions for the distortion model. In contrast, we provide an image simulated by the Fredholm Integral model with shift-invariant distortion functions. Fig. \ref{fig_example} shows the results of the simulations.

Simulations in Fig. \ref{fig_example} indicates that the Fredholm Integral model meets the theoretical expectations for field distortion. PSFs in the shift-invariant simulations are identical regardless of their source's position. In simulations with non-trivial distortion functions, the sources' position influences their corresponding PSF's shape and position. 

\begin{figure*}[!t]
	\centering
	\includegraphics[scale = 0.25]{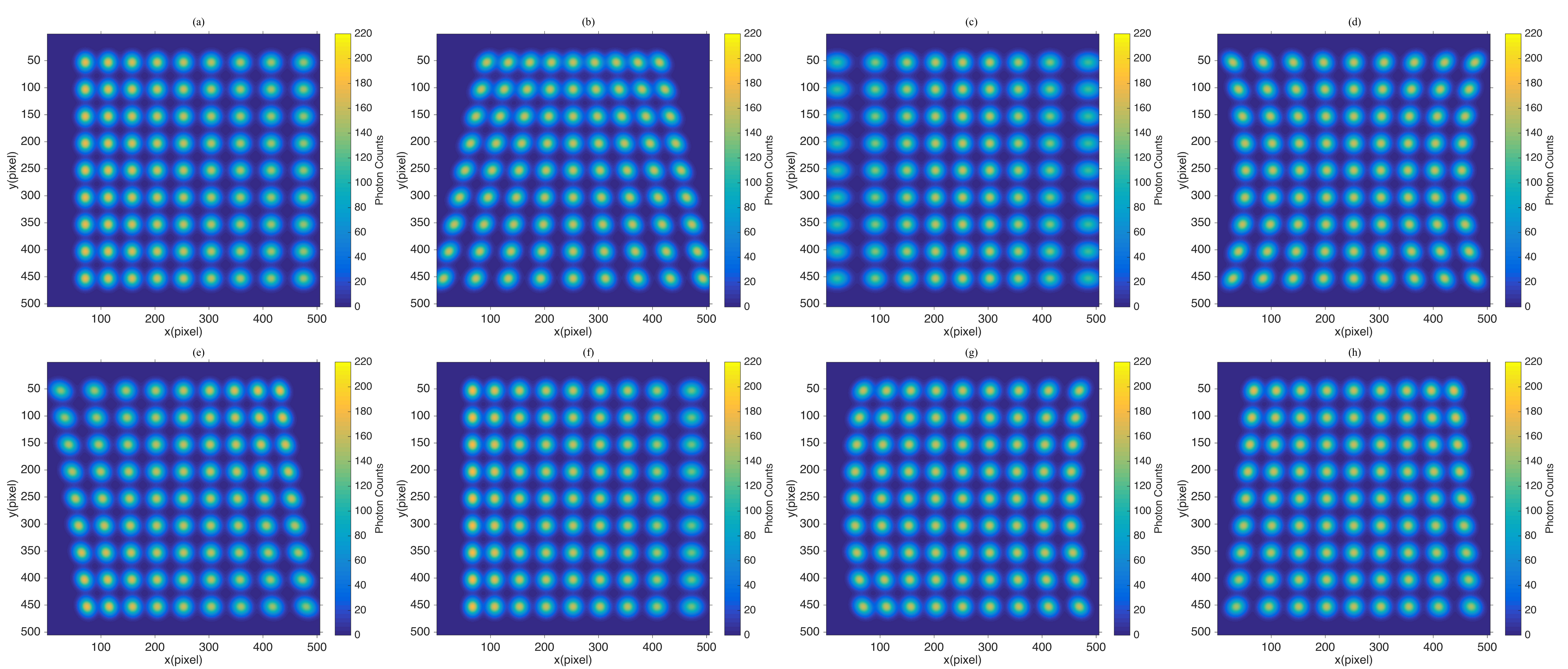}
	\caption{Simulations of single polynomial distortion functions. The distortion kernel is given by $K_d = \exp(-\frac{1}{2\sigma^2}((u-x+f)^2+(v-y)^2))$, where $\sigma = 10$. Single polynomial functions and their corresponding parameters are listed in table \ref{table_SPP}. Simulations generated by single polynomial terms in distortion function $g$ are omitted because of the symmetry in coordinates.}
	\label{fig_SPP}
\end{figure*}

Since we concern mostly the polynomial distortion functions, simulations of single polynomial distortion functions are also included in Fig. \ref{fig_SPP}. Typical distortion patterns like pincushion distortion and trapezoidal distortion are included in Fig. \ref{fig_SPP}b, d and h. While Fig. \ref{fig_SPP}a, c, e, f and g show more distortion patterns.

\begin{table*}[!t]
	\caption{Single Polynomial Functions and Corresponding Parameters}
	\label{table_SPP}
	\centering
		\begin{tabular}{|c|c|c|c|c|c|c|c|c|}
		\hline
		Single Polynomial Function &$ux$ &$vx$ &$u^2x$ &$v^2x$ &$uvx$ &$u^2x^2$ &$v^2x^2$ &$uvx^2$\\
        \hline
        Parameter &5e-4 &1e-3 &3e-6 & 3e-6&3e-6 &1e-8 &1e-8 &1e-8\\
		\hline
	\end{tabular}
\end{table*}

Simulations of image distortions demonstrate the effectiveness of the Fredholm Integral model. This model can generate the most common distortions, including pincushion distortion and trapezoidal distortion. It also provides a new method for altering and fixing the general field distortion by offering different distortion functions, as is shown in the simulations.

\section{Methodology of measuring the distortion with Fredholm Integral model}
\label{sec3}
We propose a 2-step method to measure the field distortion for a given imaging system. This method contains the correction step for the image sensor's sampling effect and the Fredholm Integral model's measurement step. For the correction step of the image sensors' sampling effect, we discuss the sampling process's influence on the observation image and build a practical model to correct the sampling effect. For measurement step of Fredholm Integral model, We first acquire a high-resolution PSF as the reference PSF. Then we formulate an optimization problem to calculate the distortion functions' polynomial parameters. In the optimization problem, we calculate the $\chi^2$ value of the residual image to ensure the distortion parameters' proper fitting.

\label{sec3}

\subsection{Sampling effect of the image sensors}
\label{subsec3_1}

A discrete image recorded by an image sensor differs from a discrete impulse sample of the object. These two sampling processes usually are considered equivalent when the size of the image sensor is negligible. This slight difference has been noticed in discussions of varies areas including ADC analysis\cite{johansson1998time}\cite{brannon2000aperture}, image profilometry\cite{chen1999error}\cite{chen2009error}, and image processing\cite{2001Effect}\cite{2012Error}. Researchers usually treat the difference between the two sampling method as an intrinsic error. However, the high accuracy field reconstruction concerns the sub-pixel accuracy measurement, and the difference between the two sampling process need to be corrected. We show the difference between an ideal discrete sample image and an image sampled by an image sensor with a fundamental matrix model for the sampling process and provide a convenient method for the correction of the sampling effect.

We first make some assumptions for discussing the matrix model of the sampling process. Notice that an imaging system is a low-pass filter; the ideal image $I(u,v)$ is band-limited before recorded on the image sensor \cite{Zhai_2011}. We assume that the image sensor meets the Nyquist Sampling Theorem, such that the size of every pixel $d$ meets $d<1.22\lambda*f/(2D)$, where $\lambda$ is the wavelength of the light, $f$ is the focal length of the imaging system, and $D$ is the aperture of the imaging system. In the following discussion, we normalize the pixel's size to 1 without losing generality and neglect the gap between neighbouring pixels.

We introduce a critical constraint for the ideal image $I(u,v)$. The ideal image $I(u,v)$ should be a finite image with a definite border. The values close to its border should be around 0 to guarantee that the discrete sampling process is legal, and its frequency leakage is minimal.

For a $N\times N$ discrete image $I_{mn}$ sampled from $I(u,v)$ by a discrete impulse samplling process, we have $I_{mn} = I(u=m,v=n)$. And the sampled image $I_{mn}$ can be interpolated in the frequency domain by Shannon interpolation formula to expand its resolution and reconstruct the continuous image $I(u,v)$, such that,
\begin{equation}
	I(u,v) = \sum_{mn} I_{mn}\,{\rm sinc}(u-m)\,{\rm sinc}(v-n).
\end{equation}

We consider the image recorded by the image sensor $\tilde{I}_{mn}$ whose size is $N\times N$. The actual sampling process accumulates all the photons within the pixel located at $i$th column and $j$th row and gives the sum as a sample, which is
\begin{equation*}
	\tilde{I}_{ij} = \int_{i-1/2}^{i+1/2}\int_{j-1/2}^{j+1/2}I(u,v)\ud u\ud v,
\end{equation*}
where $I(u,v)$ is the ideal observed image. This equation reveals the difference between the actual sampling process from a discrete impulse sampling process. By studying the intrinsic connection between $\tilde{I}_{ij}$ and $I_{mn}$, we find a matrix expression which can describe their relation, such that,
\begin{equation}
\label{eqn_mat_relation}
	I_{mn} = \sum_k \left(\sum_j R^{-1}_{mi} \tilde{I}_{ij} \right) R^{-1}_{jn},
\end{equation}
where $R_{mi}=\int {\rm sinc}(x-i)H(m-x)dx$, $R_{jn}=\int {\rm sinc}(x-n)H(j-x)dx$. $H(x)$ is the rectangular function
\begin{equation}
	H(x) =
\left\{
	\begin{array}{lr}
	1, \ 0\leq|x|<1/2.\\
	0, \ otherwise.\\
	\end{array}
\right.
\end{equation} 

It is obvious to conclude that $R$ is symmetric. Besides, $R_{mi}$ and $R_{jn}$ are identical referred to their definition. In such case equation \ref{eqn_mat_relation} can be further simplified as,
\begin{equation}
	I = R^{-1}\tilde{I}R^{-1}.
\end{equation}
A detailed proof and discussion of this equation is attached in Appendix \ref{appendix}.

\begin{figure*}[!t]
	\centering
	\includegraphics[scale = 0.5]{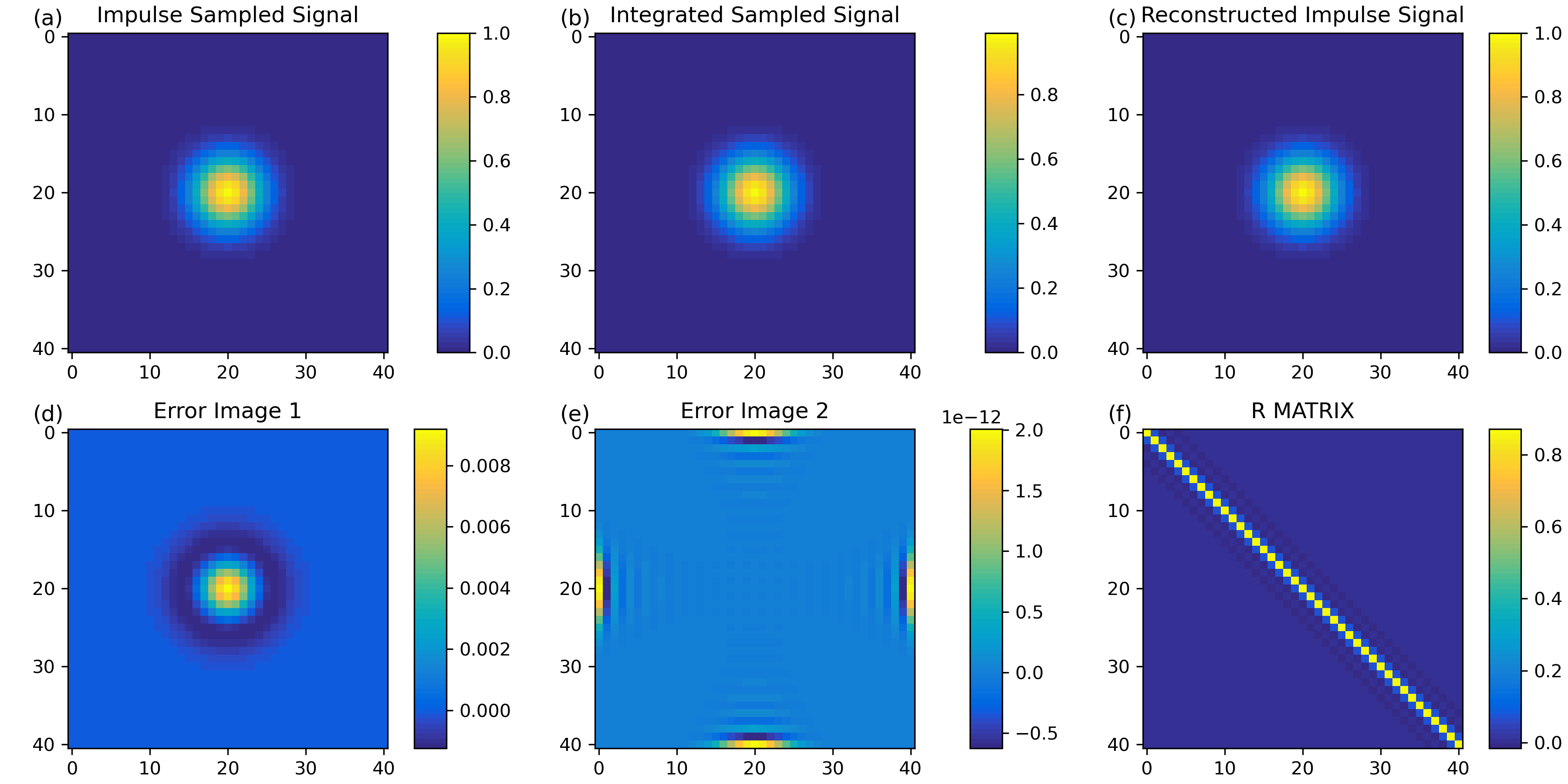}
	\caption{Simulations about sampling effect on a Gaussian spot and its correction. (a) shows the ideal discrete sampled Gaussian spot with $\sigma = 3$. (b) shows the real image sampled by an image sensor. (c) shows the corrected image from the sampling effect. (d) shows the error image of (a) and (b). (e) shows the error image of (a) and (c). (f) is the $R$ matrix.}
	\label{fig_seffect}
\end{figure*}

We show a simulation of the sampling effect on a Gaussian spot and its correction. We first generate a Gaussian spot at the centre of the field of view. We then simulate the sampled image on an image sensor and correct it with the $R$ matrix. Fig. \ref{fig_seffect} show the simulations.

In this simulation, the image sensor's sampling effect causes about 1\% deviation of the measured intensity from an ideal impulse sample image, as shown in Fig. \ref{fig_seffect}(d). By the Equation \ref{eqn_mat_relation}, we reconstruct the ideal impulse sample image from the image sampled by the image sensor, and the relative error is reduced to $2\times 10^{-12}$ at its maximum as is shown in Fig. \ref{fig_seffect}(e).

Equations and simulations that we introduced indicate the difference between a discrete impulse sampling and the image sensor's sampling can be corrected. We can reconstruct a high-accuracy discrete impulse signal $I_{mn}$ from the image sensor's sample $\tilde{I}_{mn}$ and interpolate $I_{mn}$ to recover the ideal image $I(u,v)$. In general, $I(u,v)$ has a better resolution and can reveal delicate structures hidden by the sampling effect. In the following discussion, we assume that the observed images are corrected from the sampling effect in advance.

\subsection{Measurement of the distortion with Fredholm Integral model}
The measurement step of the Fredholm Integral model contains two parts, including acquiring a high-resolution reference PSF and optimizing the distortion functions' polynomial parameters. A high-resolution PSF is a guarantee for the accuracy of measuring the distortion functions' polynomial parameters. We then add polynomial distortion functions to the high-resolution PSF and find the optimal polynomial parameters. Combining these two parts would give an optimal solution of the parameters for a given field distortion.

\begin{figure}[!t]
	\centering
	\includegraphics[scale = 0.24]{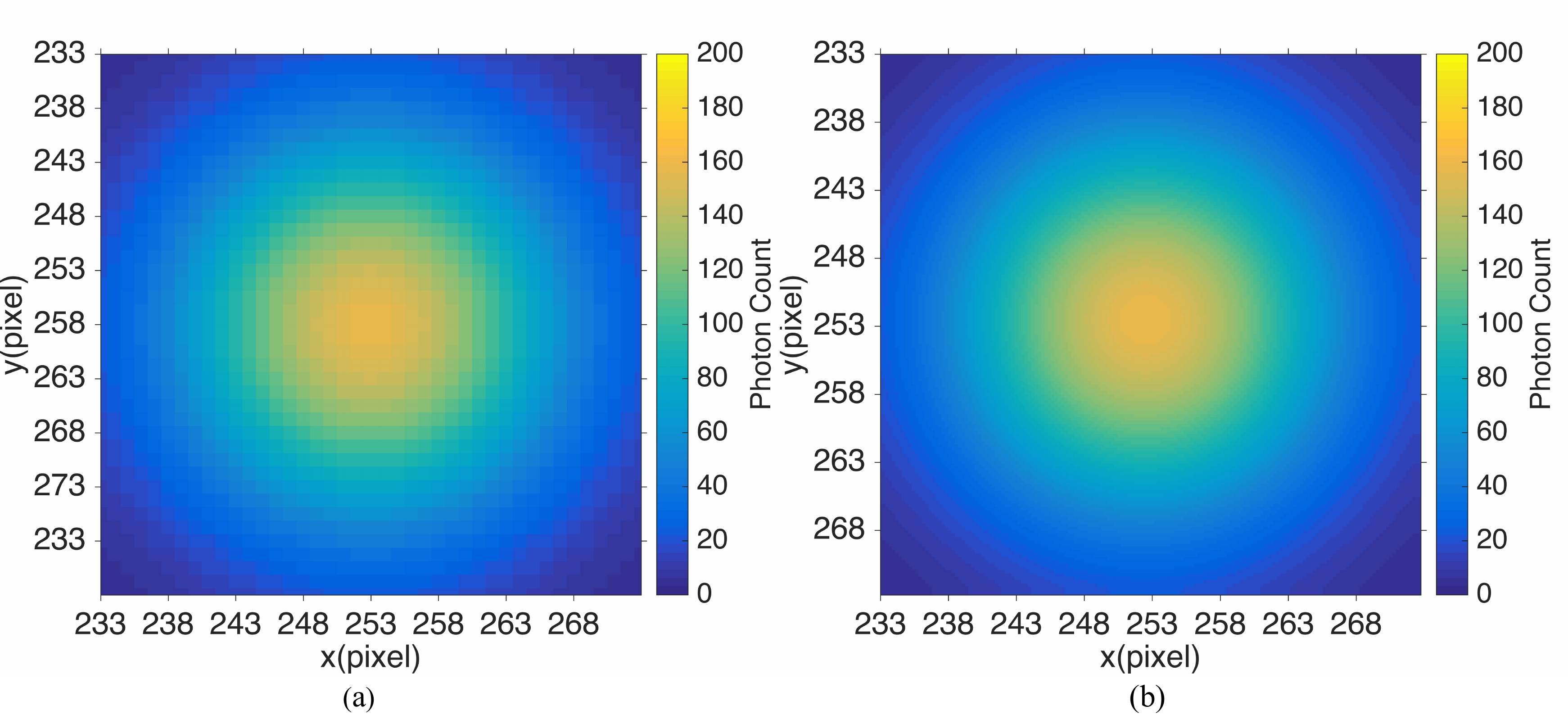}
	\caption{Reference PSF and its frequency space interpolation. (a) is the PSF extracted at the center of the field from Fig. \ref{fig_multipoly}. (b) is the high-resolution PSF after interpolation.}
	\label{fig_sim}
\end{figure}

To acquire a high-resolution reference PSF, we extract the PSF located at the centre in the field. Since the images are subject to Shannon's Sampling Theorem, we can interpolate the low-resolution PSF in the frequency space by Whittaker-Shannon interpolation formula and acquire a high-resolution one. Fig. \ref{fig_sim} shows a reference PSF extracted from Fig. \ref{fig_multipoly} and its high-resolution interpolation.

The measurement of the distortion functions' polynomial parameters in Fredholm Integal model is an optimization problem. The optimization problem is to find the parameters $\bm{\uptheta}$ whose reconstructed image best matches the $I_{mn}$. We define a value function $V(\bm{\uptheta})$ to measure the goodness of fit of the reconstructed images and $I_{mn}$, such that,
\begin{equation}
	V(\bm{\uptheta}) = \sum_{m,n} (I_{mn} - I'_{mn}(\bm{\uptheta}))^2,
\end{equation}
where $I'_{mn}(\bm{\uptheta})$ is the reconstructed distortion image generated by $\bm{\uptheta}$, $I'_{mn}(\bm{\uptheta}) = \sum_{i,j}K_{ijmn}(\bm{\uptheta})O_{ij}$, and $I_{mn}$ is the simulated distortion image. The polynomial distortion parameters $\bm{\uptheta}$ are listed in the following equation,
\begin{eqnarray*}
	f = \theta_1 ux +\theta_2 vx +\theta_3 u^2x +\theta_4 v^2x + \theta_5 ux^2 +\theta_6 vx^2,\\
	g = \theta_7 ux +\theta_8 vx +\theta_9 u^2x +\theta_{10} v^2x + \theta_{11} ux^2 +\theta_{12} vx^2.
	\label{eqn_polydef}
\end{eqnarray*}

A smaller $V(\bm{\uptheta})$ is a direct indicator of the better measurement for distortion parameters. By finding the minimum of the value function, we can determine the optimal polynomial parameters for an unknown distortion.

\section{Measurement of the distortion parameters for the simulated distortions}
\label{sec4}
Fredholm Integral model and its corresponding distortion measurement method provide a convenient algorithm for measuring the simulated distortions with optimal polynomial parameters. We validate the measurement accuracy for various simulated distortions, including polynomial distortions, distortions with Poisson noise, and distortions with relatively small PSFs. A comparison for the distortion measurement between the Fredholm Integral model and a typical pinhole distortion model is also included to illustrate the advantages of the Fredholm Integral model in high accuracy distortion measurement in multiple distortion scenes over the pinhole model. 

The implementation of the simulations and the measurement method is as follows. First, we generate several simulations with various distortion functions. Then we extract the reference PSF and define the value function $V(\bm{\uptheta})$ defined by the measurement method of Fredholm Integral model. We use a MATLAB function 'fmincon'\footnote{'fmincon' is a gradient-based nonlinear programming solver that can find a minimum of a multivariable function $f(x)$ with several constraints, including equations, inequalities, and bounds for variable $x$. Supported algorithms are interior-point optimization, sequential quadratic programming (SQP) optimization and SQP-Legacy optimization, and active-set optimization\cite{matlab}.}
 to solve the optimization problem and find the optimal parameters $\bm{\uptheta}$. We apply both interior-point optimization and SQP optimization in our method and choose the one with better results. After we measure the distortion parameters, we reconstruct the distortion images and compare them with the simulations to verify the accuracy of the distortion measurement.

\subsection{Distortion measurement for simulations with polynomial and logarithmic distortion functions}
\label{ss:simmo}
\begin{figure*}[!t]
	\centering
	\subfloat[]{\includegraphics[width=2.5in]{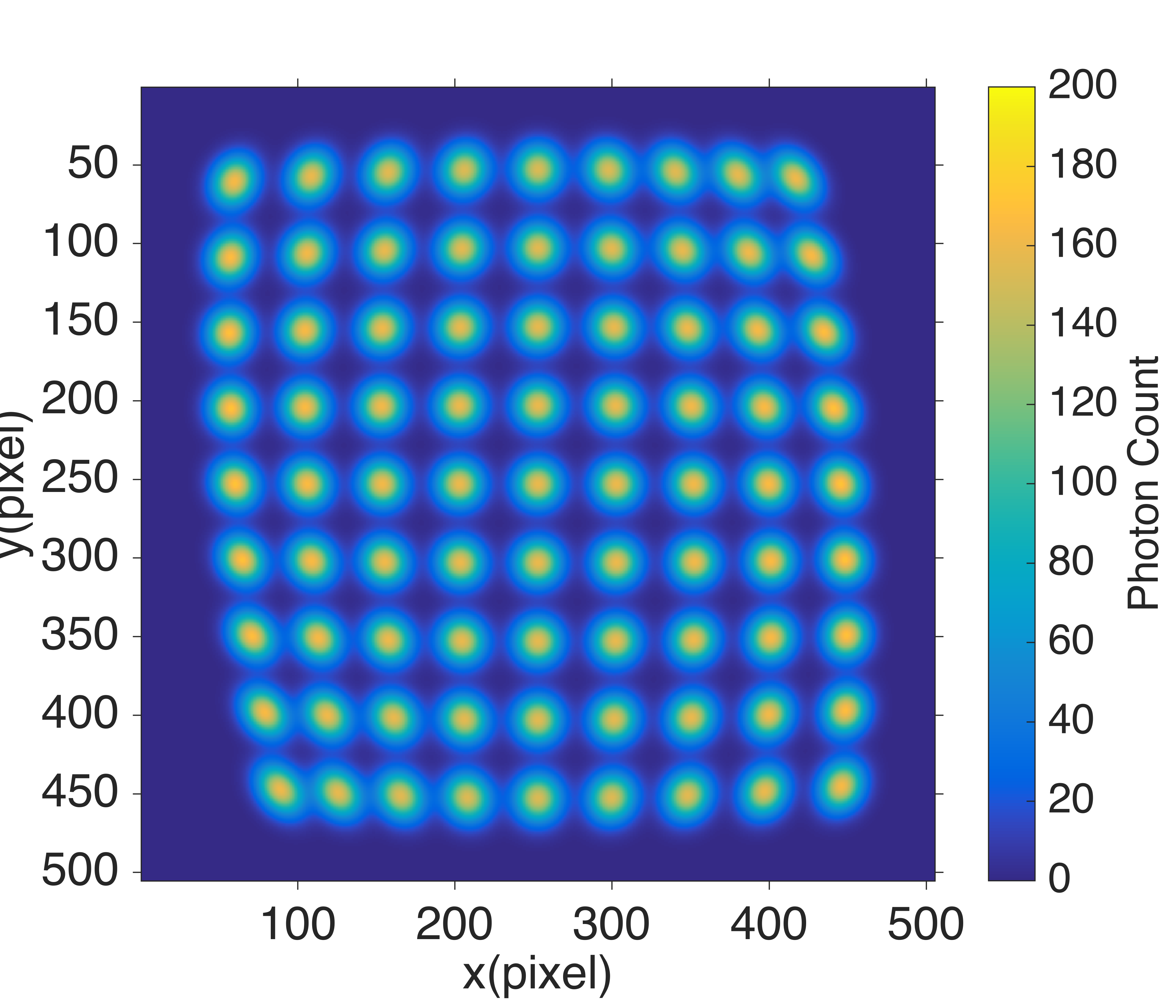}
	\label{fig_polyrec1}}
	\hfil
	\subfloat[]{\includegraphics[width=2.5in]{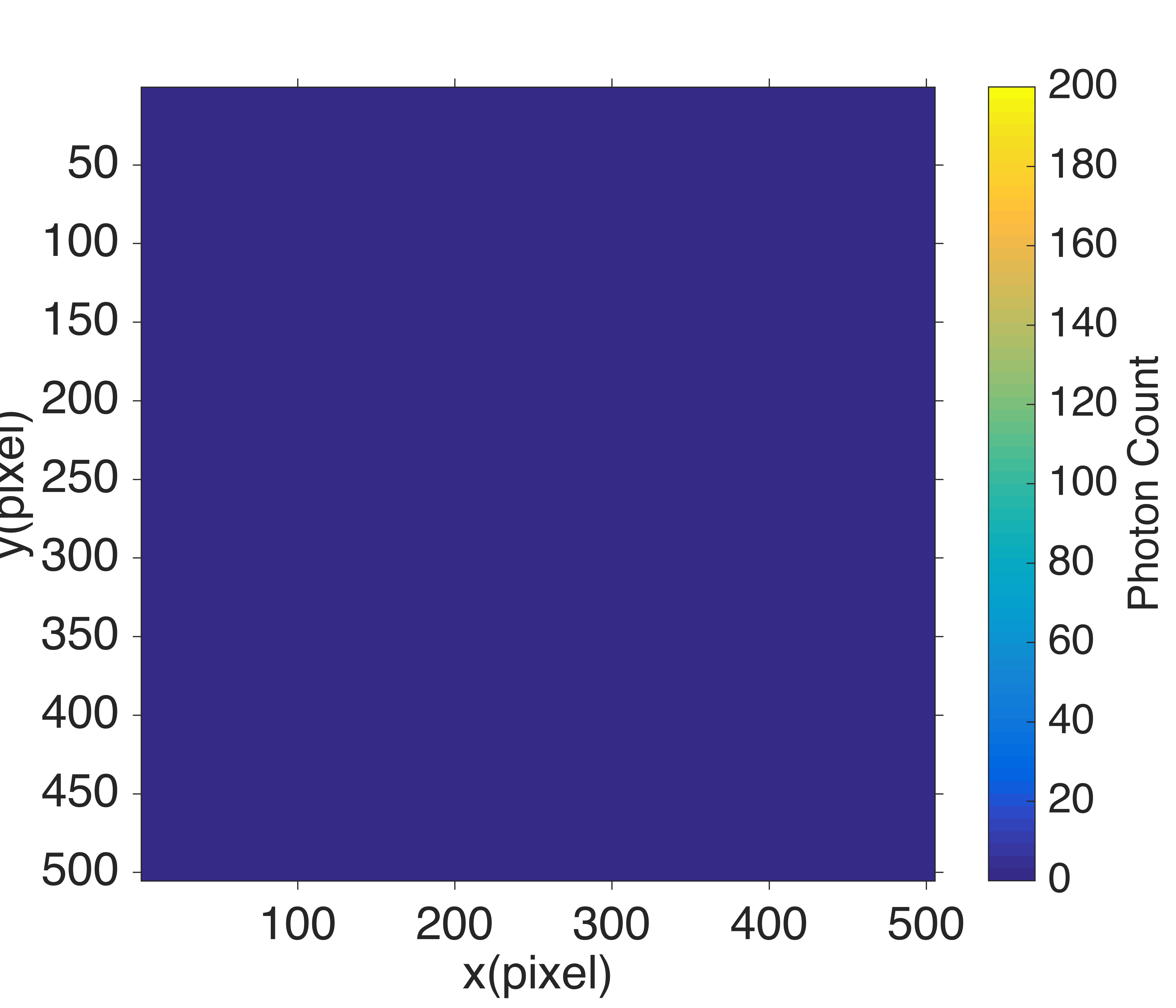}
	\label{fig_polyrec2}}
	\hfil
	\subfloat[]{\includegraphics[width=2.5in]{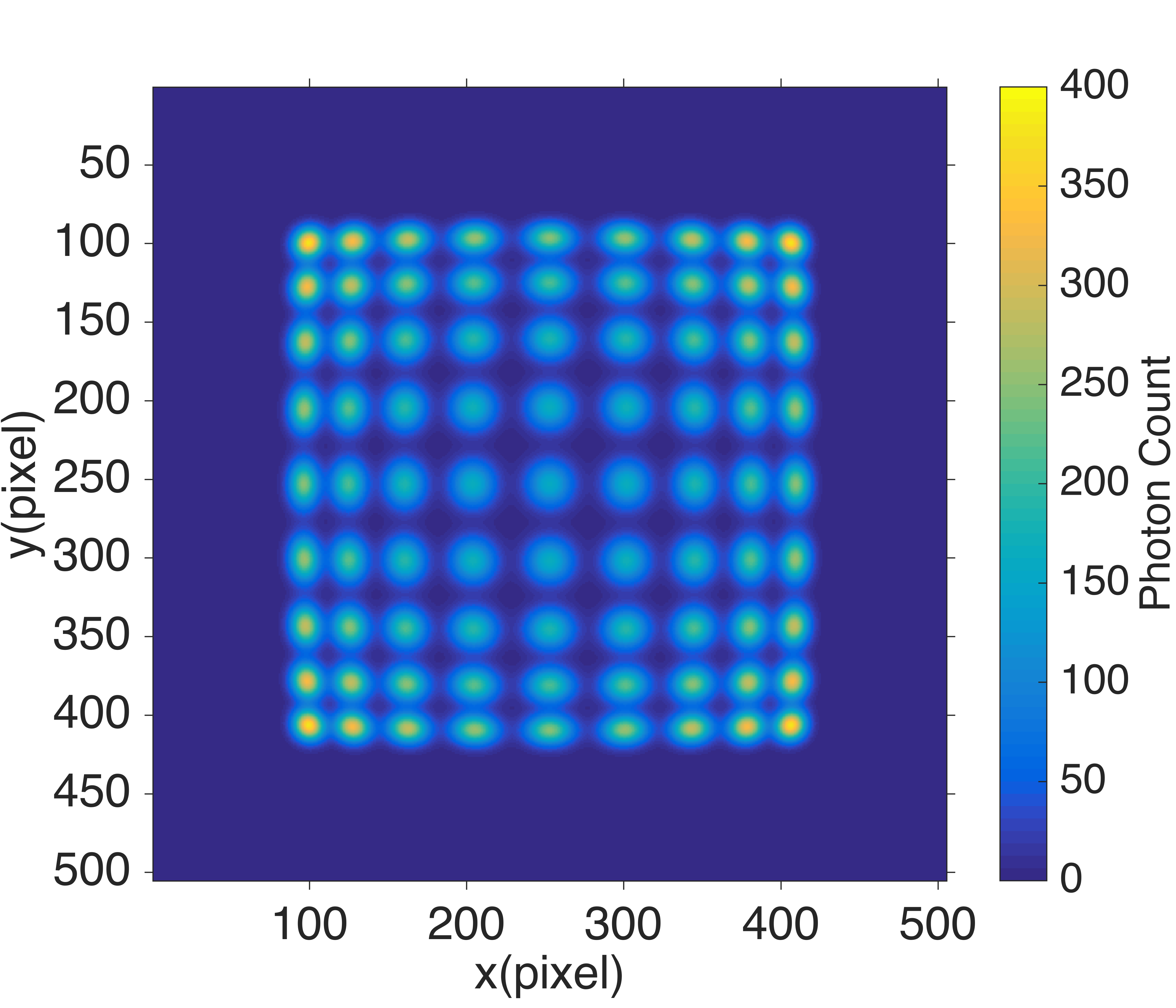}
	\label{fig_logrec1}}
	\hfil
	\subfloat[]{\includegraphics[width=2.5in]{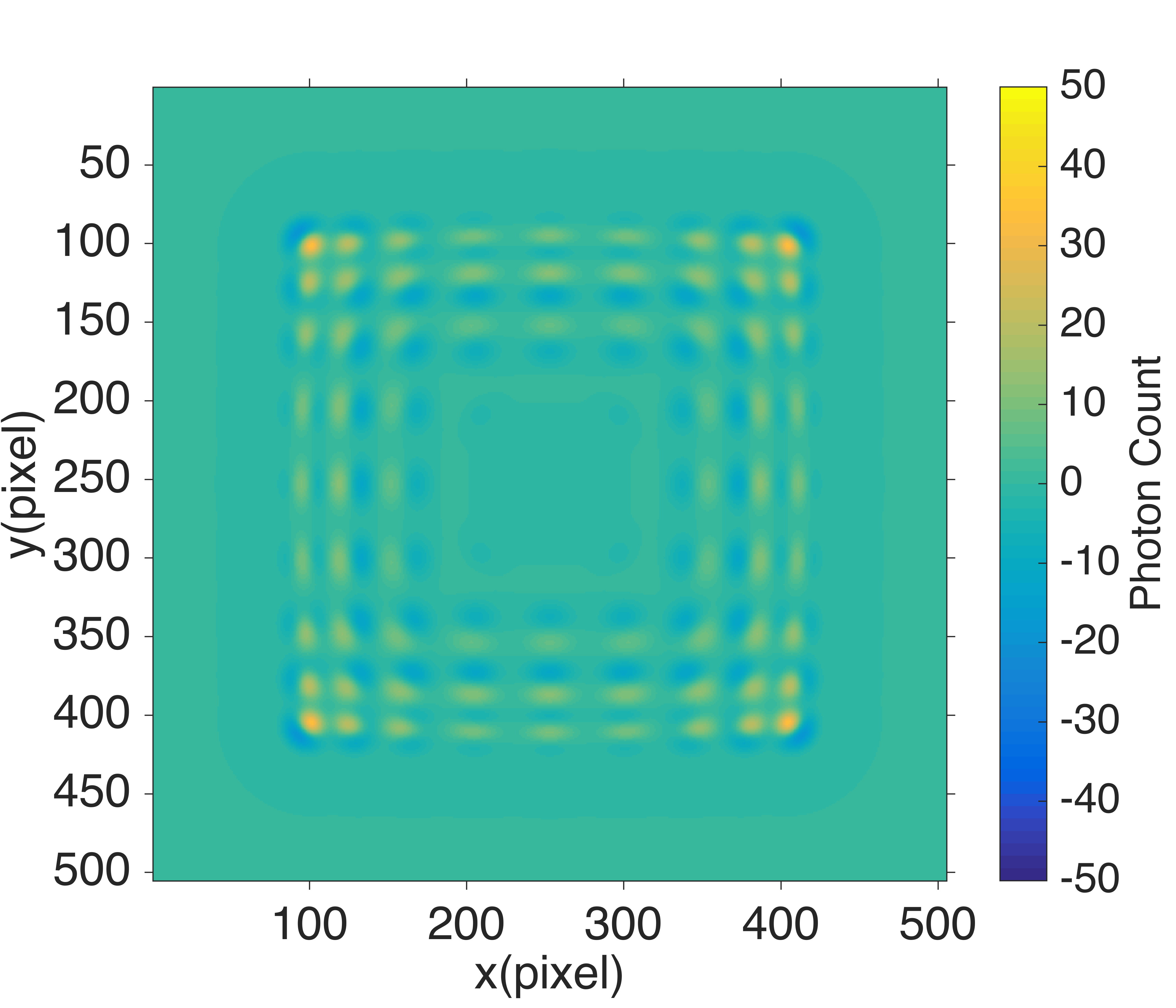}
	\label{fig_logrec2}}
	\caption{Reconstructions and residues of simulated distortions in Fig. \ref{fig_multipoly} and \ref{fig_log}. (a) and (c) are the reconstructions of polynomial distortion and logarithmic distortion respectively. (b) and (d) are residues of the reconstructed images and simulated distortions. }
	\label{fig_polyrec_A}
\end{figure*}

We apply the Fredholm Integral model and the measurement method discussed in section \ref{sec3} to measure the distortion parameters for formal simulations in section \ref{sec2}. Fig. \ref{fig_multipoly} and Fig. \ref{fig_log} are simulations with polynomial and logarithmic distortion functions. The source image is the same as Fig. \ref{fig:source}. We apply the distortion model in in section \ref{sec3} with the same definition of $\bm{\uptheta}$ and $V(\bm{\uptheta})$. The high-resolution reference PSF is a Gaussian PSF with $\sigma = 10$ . Reconstructed images with measured distortion parameters and their residues are shown in Fig. \ref{fig_polyrec_A}.

From Fig. \ref{fig_polyrec_A}, we show that there is no visible structure in Fig. \ref{fig_polyrec2}, while the residue structure of logarithmic reconstruction in Fig. \ref{fig_logrec2} is apparent. The residue structure occurs in the measurement of the logarithmic distortion because the polynomial distortion functions can not well match the logarithmic distortion functions within the boundary of the image. Both SQP optimization and interior-point optimization are repeated continuously for locating the global minimum of the optimization problem and minimizing the influence of the local minimums to the measured distortion parameters. However, local minimums still cause bad results, which is another cause for the residue structures in the Fig. \ref{fig_logrec2}.

\subsection{Distortion measurement for simulations with polynomial distortions and Poisson noise}
We apply the distortion measurement method with noisy distortions. The Poisson noise in these simulations is given by,
\begin{equation*}
	P(Noise=n) = \frac{\lambda^n}{n!}\exp(-\lambda),
\end{equation*}
where $\lambda$ indicates the average Poisson noise on single pixel. Since the simulations are normalized by the flux of sources, we can estimate the average noise $\hat{\lambda}$ with $\hat{\lambda} = \sum_{mn}{(I_{mn} - I'_{mn})}/(M\cdot N)$, where $M$ and $N$ are the size of the image, and equal $505$ in these simulations. The normalized $\chi^2$ statistic with estimated Poisson noise background and negligible deviation of source can be defined as,
\begin{equation}
	\chi^2 = \sum_{m,n} \frac{(I_{mn} - I'_{mn}(\bm{\uptheta})-\hat{\lambda})^2}{M\cdot N\cdot\hat{\lambda}}.
\end{equation} 

We use the $\chi^2$ statistic to validate these residual images with Poisson noise. If the $\chi^2$ value is around 1, we can conclude that the residual images are convincing Poisson noise, which suggests that the reconstructed parameters are a good fit of the distortion functions in our simulations. 

\begin{figure*}[!t]
	\centering
	\includegraphics[scale = 0.3]{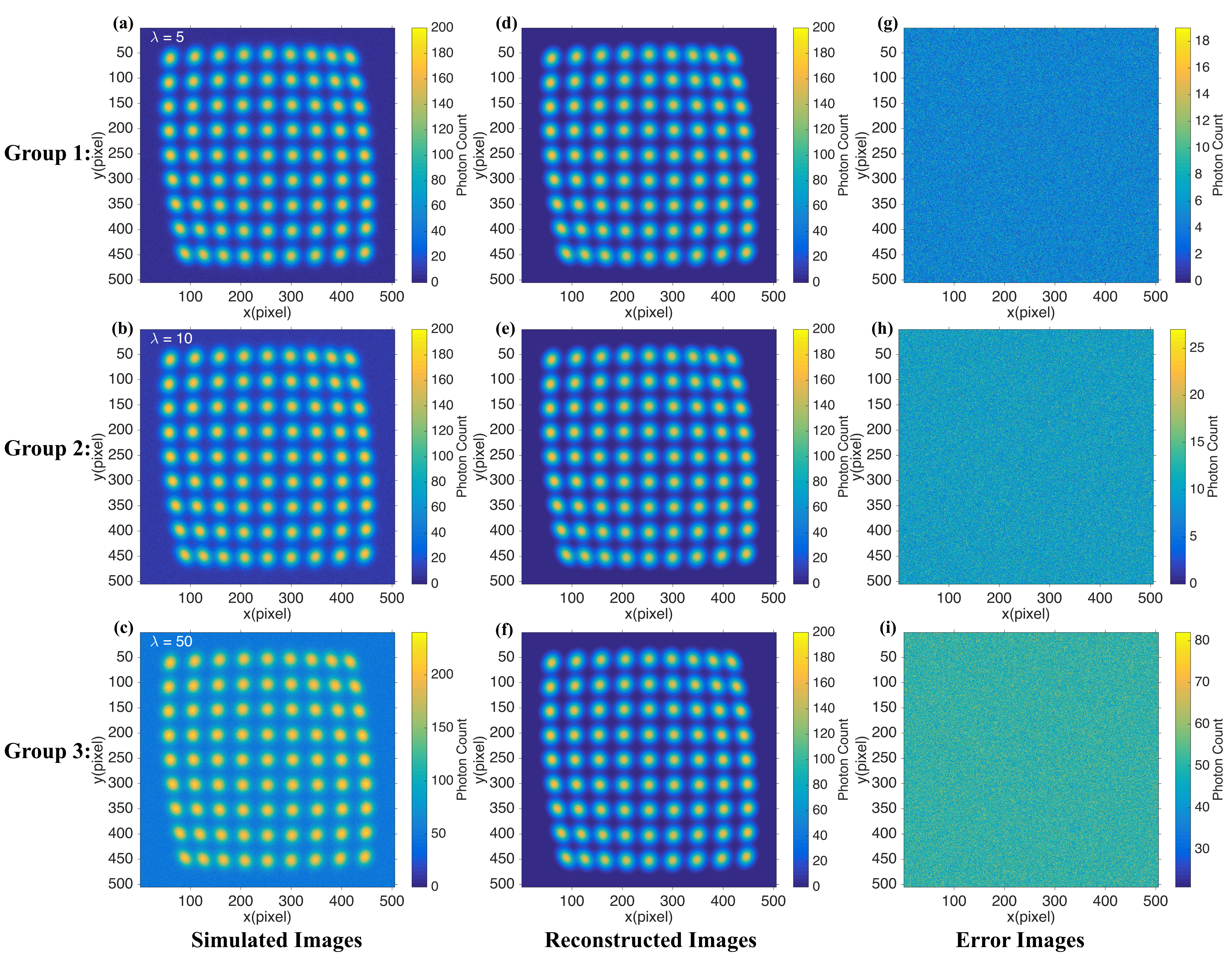}
	\caption{Reconstructions of noisy simulations with different $\lambda$. The Possion noise in (a), (b) and (c) are $\lambda = 5$, $\lambda = 10$ and $\lambda = 50$ respectively. (d), (e), (f) are reconstructed images by Fredholm Integral model. (g), (h) and (i) are residual images of the noisy images and their reconstructions.}
	\label{fig_noise}
\end{figure*}

Similarly, we define the value function $V(\bm{\uptheta})$ and solve the optimization problem with 'fmincon' tools. Together with the corresponding residue images, these noisy images and their reconstructions are divided into 3 groups by the $\lambda$ of their Poisson noise. Fig. \ref{fig_noise} shows all these images.

The polynomial parameters, their estimated Poisson noise $\hat{\lambda}$, and their corresponding $\chi^2$ are listed in the table \ref{table_RPP}.

\begin{table*}[!t]
	\caption{Results for Polynomial Parameters $\bm{\uptheta}$ Reconstruction under 3 Poisson Noise Level}
	\label{table_RPP}
	\centering
		\begin{tabular}{|c|c|c|c|c|c|c|c|c|c|c|c|c|c|c|c|}
		\hline
		\multicolumn{2}{|c|}{$\bm{\uptheta}(\times10^{-6})$} & $\theta_1$ & $\theta_2$ & $\theta_3$ & $\theta_4$ & $\theta_5$& $\theta_6$& $\theta_7$& $\theta_8$& $\theta_9$& $\theta_{10}$ & $\theta_{11}$& $\theta_{12}$& $\hat{\lambda}$& $\chi^2$\\
		\hline
		\multicolumn{2}{|c|}{Source Image} & $0$ & $-2$& $0$& $2$& $1$ & $2$& $0$& $1$& $0$ & $3$& $1$& $-1$ & & ~ \\
		\hline
		\multirow{4}*{Results} & $\lambda = 0$ & $0.00$ & $-2.00$& $0.00$& $2.00$& $1.00$ & $2.00$& $-0.00$& $1.00$& $-0.00$ & $3.00$& $1.00$& $-1.00$& & \\
		\cline{2-16}
		~ & $\lambda = 5$ & $-0.01$ & $-2.00$& $-0.01$& $1.97$& $1.01$ & $2.00$& $0.00$& $0.99$& $-0.05$ & $2.93$& $1.00$& $-1.00$& 5.0025 & $1.0076$\\
		\cline{2-16}
		~ & $\lambda = 10$ & $0.01$ & $-2.00$& $-0.12$& $1.90$& $0.99$ & $2.00$& $0.00$& $1.01$& $0.21$ & $3.05$& $1.00$& $-1.01$& 9.9968 & $1.0042$\\
		\cline{2-16}
		~ & $\lambda = 50$ & $-0.01$ & $-2.00$& $-0.17$& $1.73$& $1.01$ & $2.00$& $0.00$& $1.00$& $-0.13$ & $2.97$& $1.00$& $-1.00$& 49.9958 & $1.0084$\\
		\hline
	\end{tabular}
\end{table*}

As is mentioned before, the estimated $\hat{\lambda}$ and the $\chi^2$ are direct indicators of the goodness for the measurement. For the measured distortion parameters with Poisson noise, their estimated $\hat{\lambda}$ are close to the simulated ones. Also, the $\chi^2$ values are close to 1, indicating that the residual images are additive Poisson noise.

The results reveal that the increasing intensity of the Poisson noise might cause a slight deviation of the measured distortion parameters $\bm{\uptheta}$, while $\chi^2$ is irrelevant to the Poisson noise. This stability of $\chi^2$ illustrates that the measurement method based on the Fredholm Integral model is noise-robust and can be applied for noisy imaging systems while remaining its accuracy.

\subsection{Comparison on Fredholm Integral model and pinhole model measurement}
The traditional pinhole model is an essential one in distortion measurement and correction and has achieved a quite sufficient accuracy in most distortion measurement \cite{8678625}\cite{6365446}. While the Fredholm Integral model provides an alternative for measuring image distortion with higher accuracy. We want to compare this model with a traditional pinhole imaging model, such as a model discussed by wang\cite{wang_2008}, to examine further the Fredholm Integral model's accuracy in measuring distortion parameters. 

We briefly state a pinhole distortion model. A pinhole distortion model considers the coordinate distortion from the source image to the distorted one, such that,
\begin{equation}
\label{eqa:pinhole}
	[x_d\ y_d]^T = (1+k_1(x^2+y^2)+k_2(x^2+y^2)^2)[x\ y]^T,
\end{equation}
where $[x_d\ y_d]^T$ is the distorted coordinate on the ideal image plane from  $[x\ y]$ on the projected object plane. 

The pinhole model takes the assumption that the PSF of the imaging system should be smaller than the size of a single detector sensor. In this case, we apply a down-sample process into the Fredholm Integral model's simulation to satisfy this assumption. 

\begin{figure*}[!t]
	\centering
	\includegraphics[scale = 0.3]{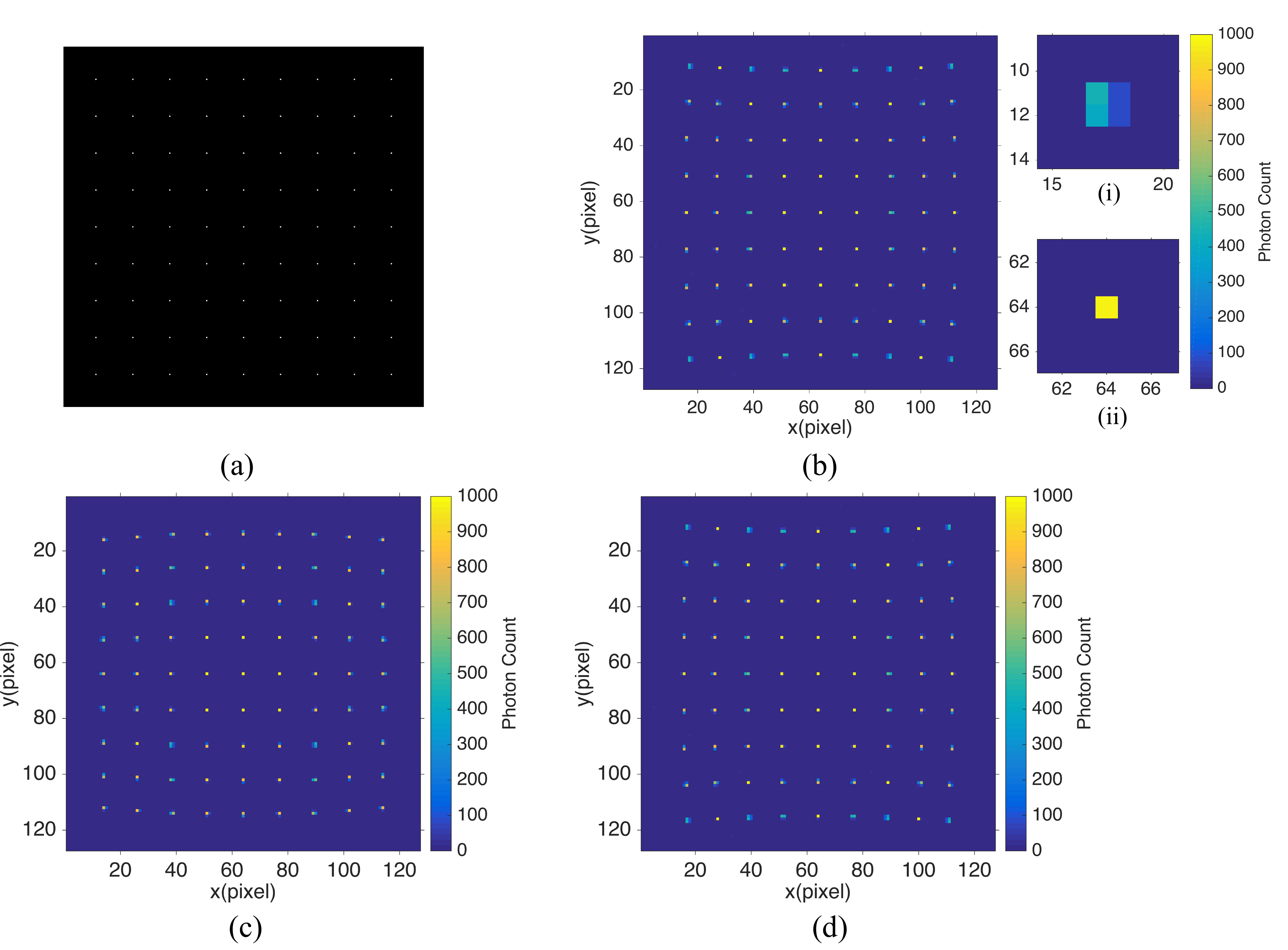}
	\caption{Source image and down-sampled distortion simulations. (a) is a high-resolution source image with $3175\times 3175$ pixels with evenly distributed point sources. The flux of a single source is 1$\times 10^3$. The reference PSF has a Gaussian shape with $\sigma = 5$. (b) is the down-sampled simulations with polynomial distortion functions. (i) and (ii) are the down-sampled PSFs at the border and the center of (b). (c) is  the down-sampled distortion with logarithmic distortion functions. (d) is the same down-sampled distortion as (b) and additional Poisson noise($\lambda = 0.01$).}
	\label{img_dsp}
\end{figure*}

Simulations in this section follow the same strategy as in section \ref{subsec_simu4model}. After that, we apply the down-sample process to compress the PSF in the simulations into about 1 pixel. We first generate a distorted image by the Fredholm Integral model with a Gaussian reference PSF with $\sigma = 5$. The original size of the source image and the simulated images is $3175$ pixel $ \times3175$ pixel. After we generate the distortion simulations, these simulated images are down-sampled 25 times to $127$ pixel $\times127$ pixel images. The down-sample process compresses the PSFs to several pixels. The source image and the simulated down-sampled images are shown in Fig. \ref{img_dsp}.

The parameters of the pinhole distortion model are expanded to $\bm{k} = (k_1,k_2,...,k_8)$ to increase its accuracy, such that,
\begin{eqnarray*}
\label{eqn_pinhole}
		x_d = (1+k_1x+k_2y+k_3x^2+k_4y^2)x,\\
		y_d = (1+k_5x+k_6y+k_7x^2+k_8y^2)y.
\end{eqnarray*}
The value function is defined as, 
\begin{equation*}
	V(\bm{k}) = \sum_{m,n} (I_{mn} - I'_{mn}.(\bm{k}))^2
\end{equation*}
The algorithm for reconstructing the distortion parameters $\bm{k}$ is similar with that for Fredholm Integral model. By minimizing the value function using 'fmincon' tool in MATLAB, we can solve the optimization problem and measure these distortion parameters $\bm{k}$.

\begin{figure*}[!t]
	\centering
	\includegraphics[scale = 0.25]{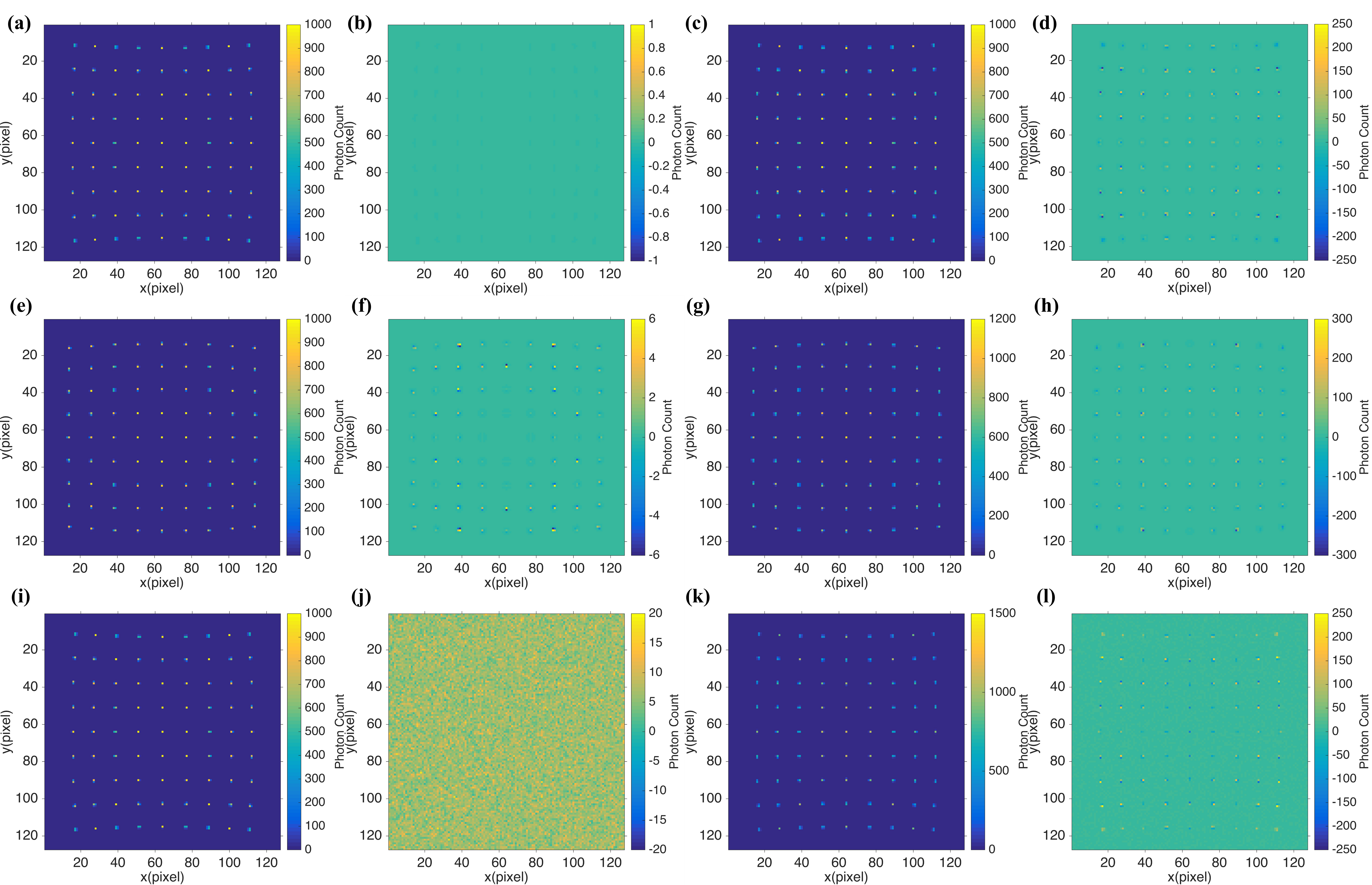}
	\caption{Comparisons on distortion reconstructions with Fredholm model and pinhole model. (a) and (c) are the reconstruction results for polynomial simulation (Fig. \ref{img_dsp}b) with Fredholm model and pinhole model respectively. (b) and (d) are the residue image of Fig. \ref{img_dsp}b and reconstructed images (a) and (c). (e) and (f) are the reconstruction results for logarithmic simulation (Fig. \ref{img_dsp}c). (f) and (h) are corresponding residual images of (e) and (g). (i) and (k) are the reconstruction results for polynomial simulation with Poisson noise (Fig. \ref{img_dsp}d). (j) and (l) are corresponding residues of (i) and (k).}
	\label{tworec}
\end{figure*}

We apply both Fredholm Integral model and pinhole model into distortion parameters' measurement and reconstruct the simulated distortions. The results are shown in Fig. \ref{tworec}.

For Fig. \ref{img_dsp}b, \ref{img_dsp}c and \ref{img_dsp}d, residual images of Fredholm Integral model are better than those of pinhole models. Table \ref{table_tworec} contains the distortion parameters measured by the Fredholm Integral model and the pinhole model and their value function. The value function in Table \ref{table_tworec} are normalized by the flux of the point source.

\begin{table*}[!t]
	\caption{Measured Distortion Parameters for Simulated Distortions by Fredholm Integral Model(FIM) and Pinhole Model(PM)}
	\centering
	\begin{tabular}{|c|c|c|c|c|c|c|c|c|c|c|c|c|c|c|}
		\hline
		\multirow{4}*{Polynomial} & \multirow{2}*{FIM($\theta\times 10^{-8}$)} & $\theta_1$ & $\theta_2$ & $\theta_3$ & $\theta_4$ & $\theta_5$ & $\theta_6$ & $\theta_7$ & $\theta_8$ & $\theta_9$ & $\theta_{10}$ & $\theta_{11}$ & $\theta_{12}$ & $V(\bm{\uptheta})$\\
		\cline{3-15}
		~ & ~ & 0.0 & 0.0 & -1.0 & 3.0 & 5.0 & 2.0 & -0.0 & 0.0 & -4.0 & -2.0 & -2.0 & 1.0& 1.4e-31\\
		\cline{2-15}
		~ & \multirow{2}*{PM($k\times 10^{-6}$)} & $k_1$ & $k_2$ & $k_3$ & $k_4$ & $k_5$ & $k_6$ & $k_7$ & $k_8$ &  &  &  &  & $V(\bm{k})$\\
		\cline{3-15}		
		~ & ~ & 0.03 & -0.05 & -3.21 & -1.14 & 0.14 & 0.05 & 1.27 & -0.63 & & & & &1.87\\
		\hline
		\multirow{4}*{Log} & \multirow{2}*{FIM($\theta\times 10^{-8}$)} & $\theta_1$ & $\theta_2$ & $\theta_3$ & $\theta_4$ & $\theta_5$ & $\theta_6$ & $\theta_7$ & $\theta_8$ & $\theta_9$ & $\theta_{10}$ & $\theta_{11}$ & $\theta_{12}$ & $V(\bm{\uptheta})$\\
		\cline{3-15}
		~ & ~ & -0.93 & 0.00 & 0.00 & 0.00 & 2.93 & 0.99 & 0.00 & -1.79 & -0.00 & -0.00 & 2.93 & 3.81 & 2.8e-3\\
		\cline{2-15}
		~ & \multirow{2}*{PM($k\times 10^{-6}$)} & $k_1$ & $k_2$ & $k_3$ & $k_4$ & $k_5$ & $k_6$ & $k_7$ & $k_8$ &  &  &  &  & $V(\bm{k})$\\
		\cline{3-15}
		~ & ~ & -0.0 & -0.19 & -1.33 & -0.45 & -0.0 & -0.02 & -1.76 & -1.29 & & & & &1.65\\
		\hline
		\multirow{4}*{Noise} & \multirow{2}*{FIM($\theta\times 10^{-8}$)} & $\theta_1$ & $\theta_2$ & $\theta_3$ & $\theta_4$ & $\theta_5$ & $\theta_6$ & $\theta_7$ & $\theta_8$ & $\theta_9$ & $\theta_{10}$ & $\theta_{11}$ & $\theta_{12}$ & $\chi^2$\\
		\cline{3-15}
		~ & ~ & -0.01 & -0.00 & -0.74 & 3.15 & 5.01 & 2.00 & -0.00 & 0.06 & -4.32 & -2.20 & -2.00 & 0.94 & 1.0020\\
		\cline{2-15}
		~ & \multirow{2}*{PM($k\times 10^{-6}$)} & $k_1$ & $k_2$ & $k_3$ & $k_4$ & $k_5$ & $k_6$ & $k_7$ & $k_8$ &  &  &  &  & $V(\bm{k})$\\
		\cline{3-15}
		~ & ~ & 0.02 & -0.03 & -3.21 & -1.14 & 0.13 & 0.04 & 1.27 & -0.62 & & & & & 2.60\\
		\hline
	\end{tabular}
	\label{table_tworec}
\end{table*}

The comparison of the distortion parameters measured by the Fredholm Integral model and pinhole model in multiple conditions suggests that the Fredholm Integral model is more accurate than the pinhole model in measuring image distortions. For images with relatively small PSFs, the Fredholm Integral model is also better than the pinhole model in describing the distortion while remaining considerable accuracy.

Results in this section cover a wide range of imaging scenes, including distorted images with and without Poisson noise and the contrast between the Fredholm Integral model and pinhole imaging model. Fig. \ref{fig_noise} and Fig. \ref{tworec} show that the distortion parameters are good estimations for the simulated distortions with clear residual images.

The results show the Fredholm Integral model's potential in measuring the distortion parameters with high accuracy, not only for imaging systems with spreading PSFs. We have also proved that the Fredholm Integral model and its reconstruction algorithm is a trusted noise-robust one, and it can handle noisy imaging systems with high accuracy. For imaging systems with small PSFs or low sample rates, the Fredholm Integral model shows better accuracy in reconstructing distortion parameters than a pinhole model in multiple scenes.

\section{Discussion}
\label{sec5}

The Fredholm Integral model is a flexible model for describing arbitrary distortions in the imaging process. It also provides a high accuracy method for measuring unknown distortions. The most important feature of the Fredholm Integral model is that it resembles the actual imaging process and is flexible for generating shift-variant distortion patterns. The models's accuracy in the arbitrary distortion's measurement can be increased by extending the polynomial terms in the polynomial distortion functions. By measuring the parameters of the polynomial distortion parameters for an unknown distortion, we can acquire a high accuracy measurement of the imaging system, which is vital for calculating the non-distorted image. The investigation of the measurement method for distortions with complicated distortion functions and Poisson noise indicates that this model is potential for high accuracy CCD calibration and image photogrammetry. 

Future studies about Fredholm Integral model would focus on several aspects. The most important one is the correction of the distortion based on the reconstructed distortion functions. Also, the Poisson distribution of the source should be considered to further improve the model's accuracy. Furthermore, methods for measuring image distortion for general objects instead of point sources are also crucial for increasing the application of the Fredholm Integral Model.

\section{Conclusion}
\label{sec6}
In this paper, we provided a general model for an arbitrary imaging process with unknown distortion based on the Fredholm integral. We also provided high accuracy method for simulating and measuring the synthesized distortion of imaging devices. Besides, we introduced the measurement error caused by sampling effect of imaging sensors and its correction method, which could increase the accuracy of the sampled image by $10^{10}$ compared to the uncorrected one. We validated the distortion measurement method based on polynomial approximation on multiple simulated distortions. For distortions with polynomial distortion functions and logarithm distortion functions, this method can measure the distortion with  cleaner residual image than traditional pinhole imaging model. For imaging systems with Poisson noise, this method can also achieve high accuracy measurement and is robust of the noise intensity. Since high accuracy measurement is more and more vital in image processing, the Fredholm Integral Model provides a new perspective to study images with higher accuracy and is worth further research.

\appendices
\section{Correction of the Image Sensor's Sampling Effect}
\label{appendix}

The discrete image $\tilde{I}_{ij}$ sampled by an image sensor is the convolution of  continuous image $I(u,v)$ with 2D rectangular function $H(u)H(v)$ , which is described by 
\begin{equation}
\label{natural-sampling-2d}
    \tilde{I}_{ij} = \iint I(u,v) H(i-u)H(j-v)dudv,
\end{equation}
where $H(x)$ is the rectangular function
\begin{equation}
	H(x) =
\left\{
	\begin{array}{lr}
	1, \ 0\leq|x|<1/2.\\
	0, \ otherwise.\\
	\end{array}
\right.
\end{equation}

The Whittaker-Shannon interpolation formula shows the connection between the ideal sampled discrete image $I_{mn}$ and the continuous image $I(u,v)$, such that,
\begin{equation}
\label{ws-formula-2d}
	I(u,v) = \sum_{mn} I_{mn}\,{\rm sinc}(u-m)\,{\rm sinc}(v-n).
\end{equation}

Substitute $I(u,v)$ in Equation\ref{natural-sampling-2d} with Equation \ref{ws-formula-2d}, we have,
\begin{eqnarray*}
\begin{split}
\label{np-sampling-2d}
    \tilde{I}_{} = &\sum_{mn} I_{mn} \iint {\rm sinc}(u-m){\rm sinc}(v-m)\cdot \\
    					& H(i-u)H(j-v)dudv \\
                = &\sum_{mn} I_{mn} \int {\rm sinc}(u-m)\,H(i-u)du\cdot \\
                  &\int {\rm sinc}(v-m)\,H(j-v)dv \\
                = &\sum_{mn} R_{im} R_{jn} I_{mn}\\
                = &\sum_n \left(\sum_m R_{im} I_{mn} \right) R_{nj} ,
\end{split}
\end{eqnarray*}
where $R_{im}=\int {\rm sinc}(u-m)H(i-u)du$ and $R_{jn}=\int {\rm sinc}(v-n)H(j-v)dv$. It can be concluded that these matrices are equivalent and symmetric, such that$R_{im}=R_{mi} = R_{jn}$.

If we set $\tilde{\rm \bold I} = \{ \tilde{I}_{jk} \}$, ${\rm \bold I} = \{ I_{mn} \}$, ${\rm \bold R} = \{ R_{im}\} = \{ R_{jn} \}$, then the relationship between $\tilde{I}_{jk}$ and $I_{mn}$ can be represented by the following matrix equation,
\begin{equation}
\label{matrix-representation}
	\tilde{\rm \bold I} = {\rm \bold R}\, {\rm \bold I}\, {\rm \bold R}.
\end{equation}
If the inverse matrix of $\rm{\bold R}$ exist, we can calculate the ideal sample image by,
\begin{equation}
\label{inverse-matrix-representation}
	{\rm \bold I} = {\rm \bold R}^{-1}\, \tilde{\rm \bold I}\, {\rm \bold R}^{-1},
\end{equation}
which can be written out as,
\begin{equation}
\label{inverse-representation}
	I_{mn} = \sum_k \left(\sum_j R^{-1}_{mj} \tilde{I}_{jk} \right) R^{-1}_{kn}.
\end{equation}

Since we can use $\tilde{\bold{I}}$ to calculate $\bold{I}$, we can eventually reconstruct the continuous image $I(u,v)$ by Equation \ref{ws-formula-2d}.

\ifCLASSOPTIONcompsoc
  \section*{Acknowledgments}
\else
  \section*{Acknowledgment}
\fi

The authors would like to thank Dr. Xu Benda, Mr. Dou Wei and Mr. Zhang Zongyu for their kind help and suggestions for this paper.

\ifCLASSOPTIONcaptionsoff
  \newpage
\fi

\bibliography{main}

\begin{thebibliography}{10}

\bibitem{stein1997lens}
G.~P. Stein, ``Lens distortion calibration using point correspondences,'' in
  {\em Computer Vision and Pattern Recognition, 1997. Proceedings., 1997 IEEE
  Computer Society Conference on}, pp.~602--608, IEEE, 1997.

\bibitem{duane1971close}
C.~B. Duane, ``Close-range camera calibration,'' {\em Photogramm. Eng},
  vol.~37, no.~8, pp.~855--866, 1971.

\bibitem{weng1992camera}
J.~Weng, P.~Cohen, M.~Herniou, {\em et~al.}, ``Camera calibration with
  distortion models and accuracy evaluation,'' {\em IEEE Transactions on
  pattern analysis and machine intelligence}, vol.~14, no.~10, pp.~965--980,
  1992.

\bibitem{wang_2008}
J.~Wang, F.~Shi, J.~Zhang, and Y.~Liu, ``A new calibration model of camera lens
  distortion,'' {\em Pattern Recognition}, vol.~41, pp.~607--615, Feb 2008.

\bibitem{huang2016}
W.~Huang, G.~Zhang, X.~Tang, and D.~Li, ``Compensation for distortion of basic
  satellite images based on rational function model,'' {\em IEEE Journal of
  Selected Topics in Applied Earth Observations and Remote Sensing}, vol.~9,
  no.~12, pp.~5767--5775, 2016.

\bibitem{sawhney1999true}
H.~S. Sawhney and R.~Kumar, ``True multi-image alignment and its application to
  mosaicing and lens distortion correction,'' {\em IEEE Transactions on Pattern
  Analysis and Machine Intelligence}, vol.~21, no.~3, pp.~235--243, 1999.

\bibitem{sun2016camera}
Q.~Sun, X.~Wang, J.~Xu, L.~Wang, H.~Zhang, J.~Yu, T.~Su, and X.~Zhang, ``Camera
  self-calibration with lens distortion,'' {\em Optik-International Journal for
  Light and Electron Optics}, vol.~127, no.~10, pp.~4506--4513, 2016.

\bibitem{Goljan_2014}
M.~Goljan and J.~Fridrich, ``Estimation of lens distortion correction from
  single images,'' {\em Media Watermarking, Security, and Forensics 2014}, Feb
  2014.

\bibitem{fitz2001}
A.~W. Fitzgibbon, ``Simultaneous linear estimation of multiple view geometry
  and lens distortion,'' in {\em Computer Vision and Pattern Recognition, 2001.
  CVPR 2001. Proceedings of the 2001 IEEE Computer Society Conference on},
  vol.~1, pp.~I--I, IEEE, 2001.

\bibitem{Moretti_2005}
A.~Moretti, S.~Campana, T.~Mineo, P.~Romano, A.~F. Abbey, L.~Angelini,
  A.~Beardmore, W.~Burkert, D.~N. Burrows, M.~Capalbi, and et~al., ``In-flight
  calibration of the swift xrt point spread function,'' {\em UV, X-Ray, and
  Gamma-Ray Space Instrumentation for Astronomy XIV}, Aug 2005.

\bibitem{blakeslee2002automatic}
J.~P. Blakeslee, K.~R. Anderson, G.~Meurer, N.~Ben{\'\i}tez, and D.~Magee, ``An
  automatic image reduction pipeline for the advanced camera for surveys,''
  {\em arXiv preprint astro-ph/0212362}, 2002.

\bibitem{johansson1998time}
H.~O. Johansson and C.~Svensson, ``Time resolution of nmos sampling switches
  used on low-swing signals,'' {\em IEEE Journal of Solid-State Circuits},
  vol.~33, no.~2, pp.~237--245, 1998.

\bibitem{brannon2000aperture}
B.~Brannon and A.~Barlow, ``Aperture uncertainty and adc system performance,''
  {\em Applications Note AN-501. Analog Devices, Inc.(September)}, 2000.

\bibitem{chen1999error}
W.~Chen, Y.~Hu, X.~Su, and S.~Tan, ``Error caused by sampling in fourier
  transform profilometry,'' {\em Optical Engineering}, vol.~38, no.~6,
  pp.~1029--1034, 1999.

\bibitem{chen2009error}
W.~Chen, M.~Li, and X.~Su, ``Error analysis about ccd sampling in fourier
  transform profilometry,'' {\em Optik}, vol.~120, no.~13, pp.~652--657, 2009.

\bibitem{2001Effect}
H.~Wen-Sheng, H.~Qi, and Z.~Huan-Dong, ``Effect on the fourier transform
  profilometry due to the ccd's integral sampling characteristics,'' {\em
  Optical Instruments}, 2001.

\bibitem{2012Error}
Y.~He and X.~Li, ``Error analysis of gaussian spot width measured with ccd
  sensor,'' {\em Proceedings of SPIE - The International Society for Optical
  Engineering}, vol.~8417, p.~20, 2012.

\bibitem{Zhai_2011}
C.~Zhai, M.~Shao, R.~Goullioud, and B.~Nemati, ``Micro-pixel accuracy centroid
  displacement estimation and detector calibration,'' {\em Proceedings of the
  Royal Society A: Mathematical, Physical and Engineering Sciences}, vol.~467,
  pp.~3550--3569, Aug 2011.

\bibitem{matlab}
``fmincon: Find minimum of constrained nonlinear multivariable function,'' 11
  2020.

\bibitem{8678625}
M.~{Lee}, H.~{Kim}, and J.~{Paik}, ``Correction of barrel distortion in fisheye
  lens images using image-based estimation of distortion parameters,'' {\em
  IEEE Access}, vol.~7, pp.~45723--45733, 2019.

\bibitem{6365446}
P.~{Carballeira}, J.~{Cabrera}, E.~{Ekmekcioglu}, F.~{Jaureguizar}, and
  N.~{Garc{\'\i}a}, ``Analysis of pixel-mapping rounding on geometric
  distortion as a prediction for view synthesis distortion,'' in {\em 2012
  3DTV-Conference: The True Vision - Capture, Transmission and Display of 3D
  Video (3DTV-CON)}, pp.~1--4, 2012.

\end{thebibliography}

\begin{IEEEbiography}[{\includegraphics[width=1in,height=1.25in,clip,keepaspectratio]{syq.pdf}}]{Yunqi Sun}
Yunqi Sun received a Bachelor's degree in Engineering Physics from Tsinghua University in 2017. He is currently pursuing his Ph.D. degree in Astrophysics at Tsinghua University.
His research interests include image distortion correction, image super-resolution, and image reconstruction algorithms.
\end{IEEEbiography}

\begin{IEEEbiography}[{\includegraphics[width=1in,height=1.25in,clip,keepaspectratio]{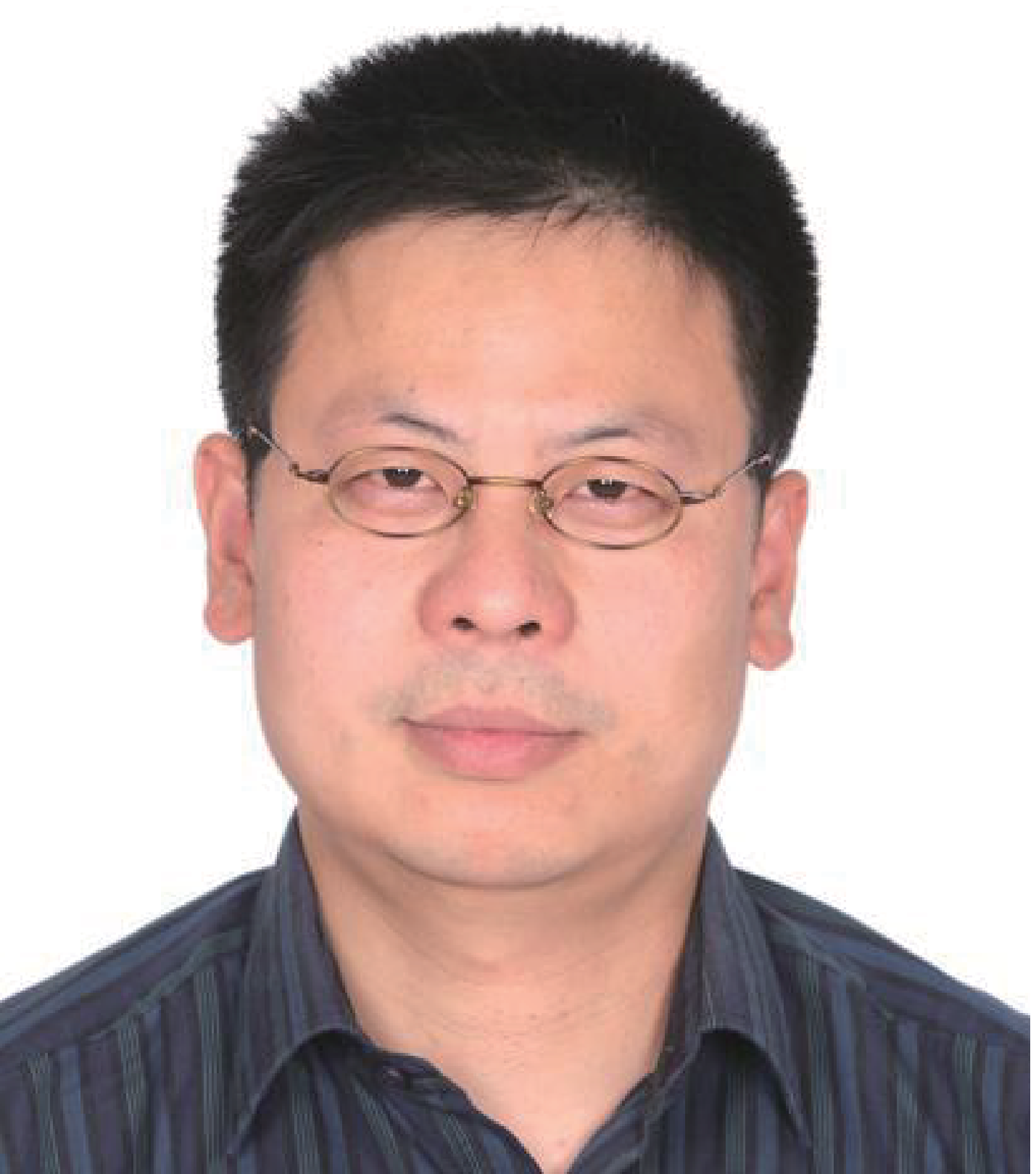}}]{Jianfeng Zhou}
Jianfeng Zhou received the B.Sc. degree in Geophysics from the University of Science and Technology of China in 1995, and the M.Sc. and Ph.D. degrees in Astrophysics from Shanghai Astronomical Observatory, Chinese Academy of Sciences, in 1998 and 2001, respectively. From 2001 to 2004, he was a post-doctoral researcher at the Center for Astrophysics at Tsinghua University in China, and then joined the faculty there. He is currently an Associate Professor.

His technical interests include imaging and image reconstruction methods and astrophysics.
\end{IEEEbiography}

\ifCLASSOPTIONcaptionsoff
  \newpage
\fi

\end{document}